\documentclass[floatfix,10pt,eqsecnum,superscriptaddress,twocolumn,showpacs,showkeys,aps,pre,final]{revtex4-1}

\usepackage{graphicx}
\usepackage{dcolumn}
\usepackage{bm}
\usepackage{epstopdf}
\usepackage{amsmath}

\begin{document}

\title{Kelvin-Helmholtz instability of the Dirac fluid of charge carriers on graphene}

\author{Rodrigo C. V. Coelho}
\email{rcvcoelho@if.ufrj.br}
\affiliation{Departamento de F\'{\i}sica dos S\'{o}lidos, Universidade Federal do Rio de Janeiro, 21941-972 Rio de Janeiro, Brazil}%
\affiliation{ ETH Z\"{u}rich, Computational Physics for Engineering Materials, Institute for Building Materials
Schafmattstrasse 6, HIF, CH-8093 Z\"{u}rich, Switzerland}

\author{Miller Mendoza}
\email{mmendoza@ethz.ch}
\affiliation{ ETH Z\"{u}rich, Computational Physics for Engineering Materials, Institute for Building Materials
Schafmattstrasse 6, HIF, CH-8093 Z\"{u}rich, Switzerland}

\author{Mauro M. Doria}
\email{mmd@if.ufrj.br}
\affiliation{Departamento de F\'{\i}sica dos S\'{o}lidos, Universidade Federal do Rio de Janeiro, 21941-972 Rio de Janeiro, Brazil}
\affiliation{Instituto de F\'{\i}sica ``Gleg Wataghin'', Universidade Estadual de Campinas, Unicamp 13083-970, Campinas, S\~ao Paulo, Brazil }

\author{Hans J. Herrmann}
\email{hans@ifb.baug.ethz.ch }
\affiliation{ ETH Z\"{u}rich, Computational Physics for Engineering Materials, Institute for Building Materials
Schafmattstrasse 6, HIF, CH-8093 Z\"{u}rich, Switzerland}
\date{\today}

\begin{abstract}
We provide numerical evidence that a Kelvin-Helmholtz instability occurs in the Dirac fluid of electrons in graphene and can be detected in current experiments. This instability appears for electrons in the viscous regime passing though a micrometer scale obstacle and affects measurements on the time scale of nanoseconds.
A possible realization with a needle shaped obstacle is proposed to produce and detect this instability by measuring the electric potential difference between contact points located before and after the obstacle. We also show that, for our setup, the Kelvin-Helmholtz instability leads to the formation of whirlpools similar to the ones reported in Bandurin, D. A., et al. \textit{Science} 351.6277 (2016): 1055-1058. To perform the simulations, we develop a new lattice Boltzmann method able to recover the full dissipation in a fluid of massless particles.
\end{abstract}


\maketitle

\section{Introduction}\label{introduction}

Graphene~\cite{novoselov04, novoselov05, RevModPhys.81.109} has caught a lot of attention due to its excellent electrical, mechanical and thermal properties, which open many possibilities for technological applications. Close to the charge neutrality point, the charge carriers in graphene show a relativistic dispersion relation making them behave effectively as a Dirac fluid of massless quasi-particles moving with the Fermi speed ($v_F \sim 10^6$ m/s), with a very low viscosity-entropy
ratio~\cite{muller09} and very high thermal conductivity~\cite{baladin08}. It also shows an extremely high electrical mobility, reaching saturation velocities above $ 3\times 10^5$ m/s for low carrier densities even at room temperature~\cite{dorgan10}.

Recently there has been a great interest in the hydrodynamic regime of charge carriers in conductors. To achieve this regime, the electron-electron scattering must dominate over the electron-impurities and the electron-phonon scattering, which is difficult to obtain for most metals and semi-conductors.
Before graphene, one of the few observations of such hydrodynamic effects in solids was an analogue of Poiseuille flow in two-dimensional high mobility wires of (Al,Ga)As heterostructures~\cite{jong1995hydrodynamic} theoretically predicted by Gurzhi~\cite{gurzhi1968hydrodynamic}. Recent experiments have shown that electrons in graphene exhibit hydrodynamic behavior for a wide range of temperatures and carrier densities~\cite{bandurin16}, due to  weak electron-phonon scattering~\cite{PhysRevLett.103.226801} and to new technologies to produce ultra-clean samples~\cite{skakalova2014graphene}. Remarkably, the formation of whirlpools (vortices) in graphene was predicted and subsequently observed~\cite{PhysRevB.92.165433, pellegrino2016electron, levitov16, bandurin16} providing unambiguous detection of the viscous regime. Those whirlpools are able to explain the observed negative resistance close to contacts. Another evidence for the hydrodynamic regime in graphene was found for electrons passing through a constriction Refs.~\cite{guo2017higher,kumar2017super}. In this experiment, the measured electrical mobility exceeds the maximum limit predicted for the ballistic regime, but can be explained by the hydrodynamic model. In addition, a signature of the Dirac fluid was pointed out in Ref.~\cite{crossno16} by the observation of a breakdown of the Wiedemann-Franz law close to the charge neutrality point.

The Kelvin-Helmholtz instability (KHI) is one of the most famous instabilities in fluid dynamics and it is an important mechanism for the formation of vortices and precursor of turbulence~\cite{smyth2012ocean, PhysRevLett.62.772, wyper2013kelvin}. It appears when two fluids, or two parts of the same fluid, are sheared against each other with a small perturbation at the interface~\cite{chandrasekhar61}. It occurs in many situations in nature, as with fluctus clouds in the sky, the waves on the beach or the red spot of Jupiter and it plays an important role to understand phenomena in magnetohydrodynamics~\cite{mohseni15} as the interaction between the solar wind and the Earth's magnetosphere~\cite{hasegawa04}. It was also observed experimentally~\cite{Blaauwgeers02} in superfluid $^3$He. The KHI does not appear for supersonic relative speeds between the two fluids~\cite{bodo04}, which explains the stable flow for relativistic planar jets in astrophysical systems as galactic nuclei and gamma-ray bursts~\cite{perucho04, 0143-0807-36-1-015007}. 

In this paper, we provide numerical evidence that the KHI can be produced and detected in current experiments on the Dirac fluid in graphene. Since most of the recent studies are on the steady states of the flow (e.g., whirlpools), our proposal to observe the KHI should make it possible to explore also  transient states, complementing our understanding about the hydrodynamic regime of electrons. We first simulate an idealized system to observe the appearance of the so-called cat-eyes pattern in the charge density field when we have shear between two regions of the fluid. Next, we simulate the fluid of electrons passing by an obstacle of micrometric scale, which creates a shear in the fluid, and analyze the impact of the KHI on the electric potential difference (EPD) between two contact points before and after the obstacle. According to our simulations, the duration of the instability is on the time scale of nanoseconds. Since this is challenging to observe experimentally, we suggest to produce it many times by using an alternating squared current of few hundreds of megahertz, and later take the statistical average of the signal. As we will see, the KHI leads to the formation of whirlpool-like regions similar to the ones in Ref.~\cite{bandurin16}.

The Boltzmann equation~\cite{cercignani02, kremer10} is widely used to derive hydrodynamic equations for graphene, since the macroscopic collective behavior of charge carriers, not always recovered by standard hydrodynamics, can be calculated from first principles~\cite{narozhny2017hydrodynamic, PhysRevB.92.115426,  PhysRevB.78.085416, PhysRevLett.92.026803, PhysRevB.78.115406, PhysRevB.85.195421, PhysRevB.91.035414, PhysRevB.93.125410}. In Ref.~\cite{PhysRevB.92.115426}, the generalized Navier-Stokes for electronic flow in graphene is derived with a procedure similar to the Chapman-Enskog expansion~\cite{chapman70}. Interestingly, the resulting hydrodynamic equations are not Lorentz or Galilean invariant due to nonlinear terms, which are specially relevant in the high velocity regime. The Boltzmann equation is not valid at the quantum critical point where charge density and temperature are equal to zero. Nevertheless in  experiments performed at finite carrier density, controlled by an external gate voltage, the Boltzmann equation is expected to give reliable results~\cite{RevModPhys.81.109}.

The Lattice Boltzmann Method (LBM)~\cite{kruger16, succi01} is a computational fluid dynamics technique based on the space-time discretization of the Boltzmann equation that has been successfully applied to simulate classical, semi-classical~\cite{coelho14,coelho16-2,yang09}, quantum~\cite{palpacelli2008quantum, PhysRevE.76.036712, solorzano2017lattice} and relativistic fluids. It has many advantages over other numerical methods as the facility to simulate flows through complex geometries and the easy implementation and parallelization of computational codes. The relativistic version of LBM~\cite{mendoza10, PhysRevE.95.053304} has been extensively used in the literature to simulate the Dirac fluid in graphene~\cite{mendoza11,oettinger13, furtmaier15, mendoza13, giordanelli2017modelling}. This approach naturally includes the linear dispersion relation and the relativistic equation of states by treating the quasi-particles in graphene as ultra-relativistic particles, analogously to models for the Quark-Gluon plasma~\cite{hupp11, romatschke11,mendoza13-3,hwa2010quark,teaney09}, which is a truly relativistic fluid. The speed of light in this approach is played by the Fermi speed and a low macroscopic velocity regime is always adopted, making the relativistic corrections disappear. The relativistic formalism is used for convenience since, the hydrodynamic equations effectively solved by these models are the standard ones~\cite{landau86}.

To perform the simulations in this paper, we develop a new relativistic LBM (considering small macroscopic velocities) for the Dirac fluid in graphene based on the expansion of the Fermi-Dirac distribution up to fifth order in orthogonal polynomials following the procedure developed in Ref.~\cite{coelho16-2}. According to the 14-moment Grad's theory, the fifth order expansion of the equilibrium distribution function (EDF) is needed to recover the full dissipation in the fluid, i.e., the Navier-Stokes equation and Fourier's law~\cite{mendoza13-3, mendoza13-2, cercignani02}, which is necessary to have an accurate description for instabilities and other viscous effects. The previous models for graphene using a similar approach were limited to a second order expansion~\cite{mendoza11,oettinger13}. 

This work is organized as follows. In Sec. \ref{model-description-sec} we describe our model, including the fifth order expansion in relativistic polynomials and the new quadrature required by this expansion. More details about the model can be found in the Supplemental Material~\footnote{See Supplemental Material at [URL will be inserted by publisher] for more details about the model, which includes the Refs.~\cite{mendoza10, kruger16, muller09, coelho14, doria17, coelho16-2, cercignani02}.}, as the full description of the polynomials, the quadrature with high precision and the explicit expansion of the EDF. Due to the novelty of our model, we first validate and characterize it in Sec. \ref{numerical-tests-sec}. The Riemann problem is performed and the solution is compared with a reference model. We find the viscosity-relaxation time relation through the Taylor-Green vortex decay and also find the thermal conductivity-relaxation time relation by analyzing the Fourier flow. In Sec. \ref{khi-sec}, the KHI for graphene is studied and an  experimental realization is proposed. In Sec. \ref{conclusions} we summarize the main findings and conclude.

\section{Model description}\label{model-description-sec}

In this section, we develop the numerical model to simulate the hydrodynamics of the Dirac fluid of charge carriers in graphene. We first review the relativistic lattice Boltzmann equation in Sec. \ref{lattice-boltzman-eq-sec}, then we expand the Fermi-Dirac (FD) up to fifth order in orthogonal polynomials in Sec. \ref{edf-expansion-sec} and, lastly, we build the Gaussian quadrature for our model in Sec. \ref{quadrature-sec}. We use the relativistic formalism to describe the relativistic dispersion relation and the equation of states of graphene. In this relativistic approach the speed of light is played by the Fermi speed. Nevertheless, the fluid moves with velocity much smaller than the Fermi speed in our setup to study the KHI. Because of this, relativistic corrections of our formalism are negligible giving the same results as standard (non-relativistic) hydrodynamics~\cite{landau86}.

\subsection{Lattice Boltzmann equation}\label{lattice-boltzman-eq-sec}

We use in our model the relativistic Boltzmann equation with the Anderson-Witting collision operator~\cite{cercignani02}, which is appropriate to treat massless particles, to describe the time evolution for the Dirac fluid:
\begin{eqnarray}\label{boltzmann-eq-general}
\bar p^\mu\partial_\mu f =   - \frac{\bar p_\mu U^\mu}{v_F^2\tau}(f - f^{eq}),
\end{eqnarray}
where $\tau$ is the relaxation time, which is a numerical parameter of our model used to tune the shear viscosity. We assume the Einstein's notation, where repeated indexes represent a sum. The greek indexes range from 0 to 2 while the latin ones range from 1 to 2. The relativistic momentum is denoted by $\bar p^\mu = (E/v_F, \mathbf{\bar p})$, 
the velocity is $U^\mu = \gamma (v_F, \mathbf{u})$ and the time-space coordinates are $x^\mu=(v_Ft, \mathbf{x})$,
where $\gamma (u) = 1/\sqrt{1-u^2/v_F^2}$ is the Lorentz factor. We use here the relativistic FD distribution,
\begin{eqnarray}\label{fd-eq-general}
f^{eq}_{FD} = \frac{1}{z^{-1}\exp\left[ \frac{\bar p_\alpha U^\alpha}{k_B T}    \right] + 1},
\end{eqnarray}
where $z=e^{\bar \mu/k_B T}$ is the fugacity. The charge carries are modeled as ultra-relativistic particles, for which the kinetic energy is much larger than the rest mass energy. Thus $\bar p^\mu \bar p_\mu =  (\bar p^0)^2 - \mathbf{\bar p}^2 =0 \:\:\Rightarrow\:\: \bar p^0 = |\mathbf{\bar p}| $, and Eq. 
\eqref{boltzmann-eq-general} becomes
\begin{eqnarray}
 \frac{\partial f}{\partial t} + \mathbf{\mathbf{v}}\cdot \nabla f = - \gamma(1  -\mathbf{ \mathbf{v}}\cdot \mathbf{u}) \frac{(f-f^{eq})}{\tau }.
 \end{eqnarray}
Here $\mathbf{v} = \mathbf{\hat{ p}} =  \mathbf{\bar p}/|\mathbf{\bar p}| $ is the microscopic velocity with norm $v_F$ and we adopt from now on natural units
$v_F=k_B=\hbar=e=1$. Note that $u/v_F \rightarrow u$ in natural units. To implement the above equation numerically, the phase space is discretized
as described in section \ref{quadrature-sec} and we use the discrete version of Eq. \eqref{boltzmann-eq-general}:
\begin{eqnarray}
&&f_\alpha(t+\delta t, \mathbf{r}+ \mathbf{\mathbf{v}}_\alpha \delta t) - f_\alpha(t,\mathbf{r}) \\ \nonumber
&&= -\gamma (1-\mathbf{\mathbf{v}}_\alpha\cdot \mathbf{u})\frac{\delta t (f_\alpha - f_{\alpha}^{eq})}{\tau},
\end{eqnarray}
where $\delta t$ is the time step of the simulations. 

In the above formalism for ultra-relativistic particles the linear dispersion relation of charge carriers in graphene was naturally included.
Nevertheless, the electronic fluid moves with a small velocity as compared to the Fermi speed ($\mathbf{u} \ll v_F \Rightarrow \gamma \approx 1$).

\subsection{Expansion of the equilibrium distribution function}\label{edf-expansion-sec}

To expand the FD distribution, we first introduce non-dimensional quantities: $\theta = T/T_0$, 
$\mathbf{p}=\mathbf{\bar p}/T_0$ and $\mu = \bar \mu/T_0$, where $T_0$ is the initial temperature. So, considering the ultra-reativistic regime, Eq. \eqref{fd-eq-general} becomes
\begin{eqnarray}\label{rfd-dist-eq}
f^{eq}_{FD} = \frac{1}{z^{-1}\exp\left[ p^0 \gamma (1-\mathbf{\mathbf{v}}\cdot \mathbf{u} ) / \theta  \right] + 1}.
\end{eqnarray}
We find the relativistic polynomials by a Gram-Schmidt procedure, with the following orthonormalization:
\begin{eqnarray}\label{ortho-eq}
&&\int \frac{d^2 p }{p^{0}}\omega(p) P^{i_1\ldots i_N}P^{j_1\ldots j_M} = \delta_{NM} \delta^{i_1\ldots i_N|j_1\ldots j_N},\nonumber \\
&&\int \frac{d^2 p }{p^{0}}\omega(p) P^{i_1\ldots i_N0}P^{j_1\ldots j_M0} = \delta_{NM} \delta^{i_1\ldots i_N|j_1\ldots j_N},\nonumber \\ 
&&\int \frac{d^2 p }{p^{0}}\omega(p) P^{i_1\ldots i_N0}P^{j_1\ldots j_M} = 0.
\end{eqnarray}
where $\omega(p)$ is the weight function, which for graphene with zero chemical potential reads:
\begin{eqnarray}\label{weight-eq}
\omega(p) = \frac{1}{e^{p}+1}.
\end{eqnarray}
Here the normalization factor is the same as for the Hermite polynomials in D-dimensions~\cite{coelho14, doria17}, where we define
$\delta_{i_1\cdots i_N\vert j_1\cdots j_N} \equiv \delta_{i_1 j_1}\cdots \delta_{i_Nj_N}\,+$ all permutations  
of  $j$'s and $\delta^{ij}$ is the Kronecker's delta. Note that we have some polynomials with only spatial components 
(latin indexes) and others which include one temporal component (zero). In principle one would have the polynomials 
$P^{\mu_1\ldots \mu_N}$ with all indexes ranging from 0 to 2, but, most of these components are zero. Below we see the polynomials for the first three orders. 
\begin{eqnarray*}
 P = A_1 
\end{eqnarray*}
\begin{eqnarray*}
 P^{i_1} = B_1 p^{i_1} 
 \end{eqnarray*}
\begin{eqnarray*}
 P^{0} = C_1 p + C_2 
 \end{eqnarray*}
\begin{eqnarray*}
 P^{i_1i_2} =  D_1 p^{i_1}p^{i_2} + (D_2 p^2 + D_3 p + D_4)\delta^{i_1i_2}
 \end{eqnarray*}
\begin{eqnarray*}
 P^{i_10} =  (E_1 p + E_2) p^{i_1}
 \end{eqnarray*}
\begin{eqnarray*}
 P^{i_1i_2i_3} &=& F_1 p^{i_1}p^{i_2}p^{i_3} + (F_2 p^2 + F_3 p + F_4) \\ \nonumber
 && \cdot (p^{i_1}\delta^{i_2i_3} + p^{i_2}\delta^{i_1i_3} + p^{i_3}\delta^{i_1i_2}) 
 \end{eqnarray*}
\begin{eqnarray*}
 P^{i_1i_20} &=&  (G_1 p + G_2) p^{i_1}p^{i_2} + \delta^{i_1i_2}(G_3 p^3 \\ \nonumber
 &&+ G_4 p^2 + G_5 p + G_6)
 \end{eqnarray*}
The fourth and fifth order polynomials are exhibited in the Supplemental Material~\cite{Note1} together with their coefficients, which can be found by applying the orthogonalization, Eq.\eqref{ortho-eq}. Notice that this tensorial form includes all possible monomials for a given dimension. Although these polynomials were derived in two spatial dimensions and for the weight function of Eq.\eqref{weight-eq}, they can also be used for other cases, as for three dimensions and for the Maxwell-J\"uttner distribution. 

The expansion of the EDF up to fifth order can be expressed as following,
\begin{eqnarray}
f^{eq} &=& \omega(\xi)\left[ \sum ^5_{N=0}\frac{1}{N!} A^{i_1\ldots i_N}P^{i_1\ldots i_N} \right.\\ \nonumber
&& \left. + \sum ^4_{M=0}\frac{1}{M!} A^{i_1\ldots i_M0}P^{i_1\ldots i_M0}   \right],
\end{eqnarray}
where $A$ are the projections of the EDF on the polynomials
\begin{eqnarray}
A^{\mu_1\,\mu_2 \cdots \mu_N}= \int \frac{d^2 \mathbf{p} }{p^0}\,f^{eq} P^{\mu_1\,\mu_2\cdots \mu_N}.
\end{eqnarray}
Notice that the denominators $N!$ and $M!$ in the expansion stem from the normalization, Eq.\eqref{ortho-eq}, as derived in Ref.~\cite{coelho16} and for the Hermite polynomials. The explicit expansion can be found in the Supplemental Material~\cite{Note1}. This expansion allows us to calculate the full set of conservation equations
for a viscous fluid and the transport coefficients, since it is required to expand up to fifth order to recover the fifth order moment of the EDF~\cite{mendoza13-3, mendoza13-2, cercignani02}:
\begin{eqnarray}\label{fifth-tensor-eq}
T_E^{\alpha \beta \gamma \delta \epsilon} = \int f^{eq} p^\alpha p^\beta p^\gamma p^\delta p^\epsilon \frac{d^2p}{p^0}.
\end{eqnarray}

\subsection{Quadrature}\label{quadrature-sec}

\begin{figure}[htb]
\center
\includegraphics[width=0.70\linewidth]{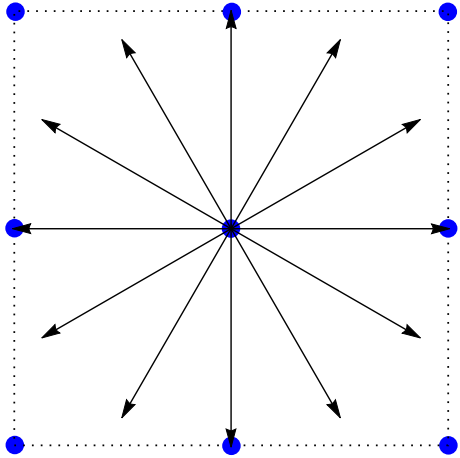}
\caption{(Color online) Velocity vectors for the D2V72 lattice. The blue points represent the spatial discretization on a square lattice.}
\label{d2v72}
\end{figure}

The Gaussian quadrature method is used to calculate numerically the integrals required to obtain the macroscopic quantities, as the charge density and the macroscopic velocity. To do so, the space is discretized by a square lattice and the microscopic velocities, with modulus $v_F$, have discrete directions. In general, to calculate the moment of order $M$,
\begin{eqnarray*}
T^{\mu_1  \ldots \mu_M} = \int \frac{d^2 p}{p^0} f^{eq} p^{\mu_1} \ldots p^{\mu_M} = \sum_{i=1}^Q f^{eq}_i p^{\mu_1}_i \ldots p^{\mu_M}_i ,
\end{eqnarray*}
we need to find the discrete weights and quadrature equations that satisfy the quadrature equation,
\begin{eqnarray}\label{quadrature-eq}
\int \frac{d^2 p}{p^0} \omega(p) p^{\mu_1}  \ldots p^{\mu_N} = \sum_{i=1}^Q w_i p^{\mu_1}_i \ldots p^{\mu_N}_i,
\end{eqnarray}
up to order $N=2M$ (in our model $N=10$) for all combinations of indexes. Because all quasi-particles move with the Fermi speed, which was considered unitary in natural units, the quadrature 
we use has 12 unitary velocity vectors, $\mathbf{v_i} = \mathbf{p_i}/|\mathbf{p_i}|$, equally distributed in the angular space, $\phi_i =  i \pi/6$ for $i=0,1,\ldots, 11$, and 72 momentum vectors (6 for each velocity vector), see Fig.\ref{d2v72}. 

We calculate the weights and momentum vectors by using the weight function of Eq.\eqref{weight-eq},
\begin{align*}
 p_1 &= 0.2520   & w_1 &= 1.4654 \times 10^{-1} \\ 
p_2 &= 1.2843  &  w_2 &= 1.6066 \times 10^{-1}  \\
p_3 &= 3.1030  & w_3 &= 5.0699 \times 10^{-2} \\
 p_4 &= 5.8738  &  w_4 &= 4.9049 \times 10^{-3}  \\
p_5 &= 9.9296  &  w_5 &= 1.2453 \times 10^{-4} \\
p_6 &=  16.0724  &  w_6 &= 4.3009 \times 10^{-7}       
\end{align*}
For higher precision, see the Supplemental Material~\cite{Note1}. Since some of the velocity vectors stream to off-lattice points, we apply a bilinear interpolation to find the populations at the lattice points. The main effect of the interpolation is to increase the effective viscosity of the fluid, what will be measured in section \ref{tg-sec}.

The Landau-Lifshitz is used to calculate the macroscopic fields from the distribution functions~\cite{cercignani02}. We first solve the eigenvalue problem 
\begin{eqnarray}\label{eigen-eq}
{T_E^{\alpha}}_\beta U^\beta = {T^{\alpha}}_\beta U^\beta =  \varepsilon U^\alpha
\end{eqnarray}
to find the energy density $\varepsilon$ and the macroscopic velocity, where the letter $E$ indicates an equilibrium field and the energy-momentum vector is calculated by
\begin{eqnarray}
T^{\mu\nu} = \sum_{i=1}^Qf_i p^\mu_i p^\nu_i.
\end{eqnarray}
Then the charge density is found by contracting the macroscopic velocity with the charge flux $N^\mu$,
\begin{eqnarray}\label{density-eq}
n =  U_\mu N_E^\mu = U_\mu N^\mu = U_\mu\sum_{i=1}^Qf_i p^\mu_i
\end{eqnarray}
Finally, we calculate the temperature by
\begin{eqnarray}\label{theta-eq}
\theta = \frac{1}{2} \frac{g_2(z)}{g_3(z)}\left(\frac{\varepsilon}{n}\right),
\end{eqnarray}
where the Fermi-Dirac integral is defined as 
\begin{eqnarray}
g_\nu(z) =\frac{1}{\Gamma(\nu)}\int_0^\infty \frac{x^{\nu-1}dx}{z^{-1}e^x+1}.
\end{eqnarray}
The chemical potential is zero in the simulations since we are considering the system close to the charge neutrality point. We calculate the temperature in Eq.\eqref{theta-eq} by using the charge density and energy density obtained with the equilibrium distribution,
\begin{eqnarray}
n = 2\pi\theta^2 g_2(z),\:\: \mbox{and}\:\: \varepsilon = 2P = 4 \pi \theta^3 g_3(z),
\end{eqnarray}
which, by Eqs.\eqref{eigen-eq} and \eqref{density-eq}, are the same for the non-equilibrium one. Here we used the equation of state ($\varepsilon=2P$, where $P$ is the hydrostatic pressure) for ultra-relativistic fluids, which is the same for Dirac fluid in graphene. The equality between the equilibrium and non-equilibrium tensors in Eqs.\eqref{eigen-eq} and \eqref{density-eq} is required to obtain the conservation of charge flow,
\begin{eqnarray}
\partial _\mu N^\mu = 0,
\end{eqnarray}
and the conservation of the energy-momentum tensor, 
\begin{eqnarray}
\partial _\mu T^{\mu \nu} =0,
\end{eqnarray}
from Eq.\eqref{boltzmann-eq-general}. However, to obtain the full dissipation, one also needs an equation for the third order non-equilibrium tensor~\cite{mendoza13-3}, which requires the fifth order equilibrium tensor, Eq.\eqref{fifth-tensor-eq}. In the Landau-Lifshitz decomposition, the charge flow can also be written as~\cite{cercignani02}
\begin{eqnarray}\label{charge-flow-eq}
N^\mu = n U^\mu - \frac{q^\mu}{h_E},
\end{eqnarray}
where $q^\mu$ is the heat flux (see Eq.\eqref{heat-flux-eq}) and $h_E = (\varepsilon + P)/n = 3\,T\,g_3(z)/g_2(z)$ is the enthalpy per particle, and the energy-momentum tensor is written as
\begin{eqnarray}
T^{\mu\nu} = p^{\langle \mu \nu  \rangle} - (P + \varpi)\Delta^{\mu\nu} + \frac{\varepsilon}{v_F} U^\mu U^\nu,
\end{eqnarray}
where 
\begin{eqnarray}
p^{\langle \mu \nu \rangle} = 2\eta \left [ \frac{1}{2}(\Delta^\mu_\gamma \Delta^\nu_\delta + \Delta^\mu _\delta \Delta ^\nu _\gamma ) - \frac{1}{3}\Delta ^{\mu\nu} \Delta_{\gamma \delta} \right] \nabla^\gamma U^{\delta}\nonumber \\
\end{eqnarray}
is the pressure deviator, $\varpi = - \mu_b \nabla_\alpha U^\alpha$ is the dynamic pressure, $\eta$ is the shear viscosity and $\mu_b$ is the bulk viscosity, which is zero for graphene~\cite{muller09}. Here, $\Delta^{\mu \nu} = \eta^{\mu \nu} - U^\mu U^\nu/v_F^2$ stands for the projector into the space perpendicular to $U^\mu$ and $\nabla^\mu = \Delta^{\mu \nu} \partial_\nu$ for the gradient operator. Note that the pressure deviator, which contains the viscosity, can only be fully recovered with a fifth order expansion.

\section{Numerical tests}\label{numerical-tests-sec}

Due to the novelty of our model, we perform three standard numerical tests, known as the Riemann problem, the Taylor-Green vortex and the Fourier flow, before applying it to the investigation of KHI in graphene. The successful comparison with reference solutions from the literature validates and characterize the present numerical procedure. 

\subsection{Riemann problem}

\begin{figure}[h]
\includegraphics[width= \linewidth]{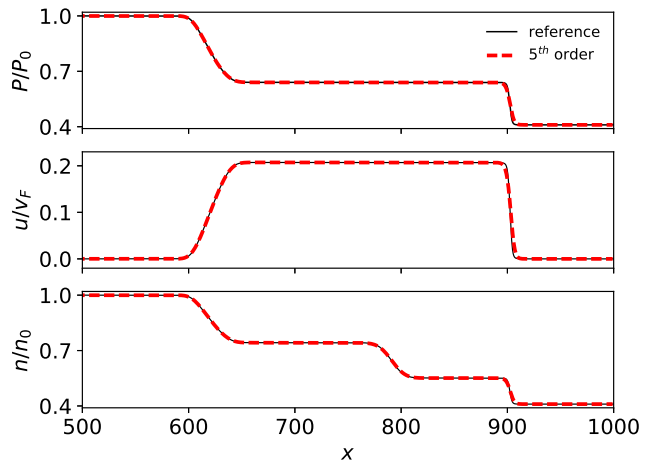}
\caption{ (Color online) Pressure, velocity and charge density for the Riemann problem comparing our model and the reference model for two dimensions~\cite{mendoza13-3}.}
\label{shock-tube-fig}
\end{figure}

We validate our code by performing the Riemann problem, which is a benchmark test for fluid dynamical models, and we compare with the result from the model described in Ref.~\cite{mendoza13-3}. In the Riemann problem, two regions of the fluid, with different states (for instance, with different velocities or densities), are separated creating a discontinuity and, then, the system evolves forming compression and rarefaction shock waves. For the simulations
we use a constant relaxation time $\tau=0.9$, an effectively one dimensional system of size $L_x\times L_y=1000\times2$ nodes and periodic boundary conditions in both directions. Initially, we have $\mathbf{u}=0$ and $\theta = 1$ everywhere and the density is 
$n_0=1$ at $L_x/4<x<3L_x/4$ and $n_0=0.41$ elsewhere. In Fig. \ref{shock-tube-fig} we see the results 
after 200 time steps, which are in excellent agreement with the reference model (adapted for two spatial dimensions) described in Ref.~\cite{mendoza13-3}. Only half 
of the space is shown ($L_x/2<x<L_x$) since the other part is an exact reflection of this one.

\subsection{Taylor Green vortex}\label{tg-sec}

\begin{figure}[htb]
\center
\includegraphics[width= \linewidth]{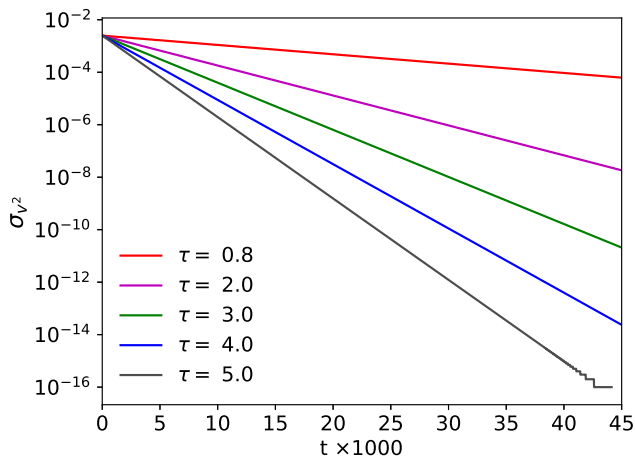}
\caption{(Color online) Standard deviation of the squared velocity as a function of time in the Taylor-Green vortex decay for 
5 different relaxation times using our 5$^{th}$ order model.}
\label{log-plot-fig}
\end{figure}

\begin{figure}[htb]
\center
\includegraphics[width= \linewidth]{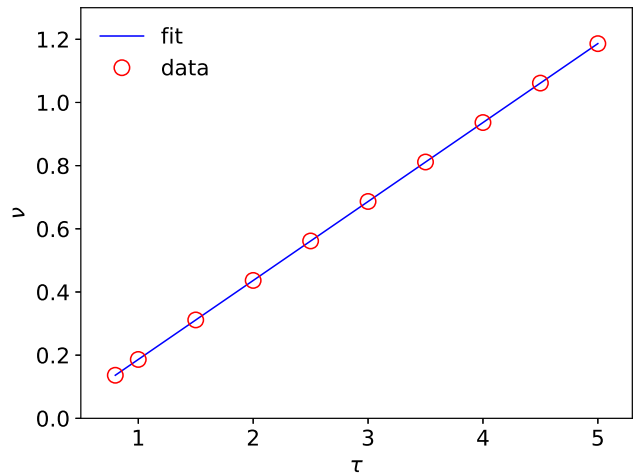}
\caption{(Color online) Relation between the kinematic viscosity and the relaxation time obtained with the Taylor-Green vortex.}
\label{eta-tau-fig}
\end{figure}

The Taylor-Green vortex decay is a numerical experiment to measure the viscosity of a fluid and it consists of initializing four vortices and analyzing their decays with time. For this problem the Navier-Stokes equations can be solved exactly for low velocities, which gives an exponential decay in time of the kinetic energy with rate depending on the kinematic viscosity $\nu$, $\mathbf{u}(x,y,t) = \mathbf{u}_{0}(x,y)e^{-2\nu t (2\pi/L)^2}$, where $\mathbf{u}_0$ is the initial velocity and $L$ the length of the squared domain~\cite{mei06}. We simulate a system of size $L\equiv L_x=L_y=512$ nodes for 10 different relaxation times, ranging from 0.8 to 5.0 for 45000 time steps. The initial conditions are $n_0=1.0$ and $\theta _0=1.0$ in the whole domain and the initial velocities are:
\begin{eqnarray}
&&u_{0x} (x,y) = - u_0  \cos\left(\frac{2\pi x}{L}\right)  \sin\left(\frac{2\pi y}{L}\right)   \\ 
&&u_{0y} (x,y) =  u_0  \sin\left(\frac{2 \pi x}{L}\right)  \cos\left(\frac{2 \pi  y}{L}\right)  , 
\end{eqnarray}
where $u_0=0.1$. We also set the initial non-equilibrium distribution as described in Ref. ~\cite{mei06} in order to reduce
the oscillations in kinetic energy. So the average squared velocity is
\begin{eqnarray}
\langle u^2 \rangle = \int_0^{L}\int_0^{L} \frac{dx dy}{L^2}(u_x^2+u_y^2) = \frac{u_0^2}{2}e^{-16\pi^2\nu t/L^2} 
\end{eqnarray}
and the standard deviation for $u^2$ is
\begin{eqnarray}\label{sigma-u2}
\sigma_{u^2} = \sqrt{\int _0 ^L \int _0^L \frac{dx dy}{L^2} (u^2 - \langle u^2 \rangle )^2}  =\frac{u_0^2}{4} e^{-16\pi^2\nu t/L^2} .\nonumber \\
\end{eqnarray}
In Fig. \ref{log-plot-fig} we see $ \sigma_{u^2} $ as a function of time in semi-log scale. By Eq. \eqref{sigma-u2}
the slope of $ \sigma_{u^2}(t)$ is $(-16\pi^2\nu /L^2)$, which allows us to measure the kinematic viscosity $\nu$.
Fig. \ref{eta-tau-fig} shows the kinematic viscosity as a function of the relaxation time.
The theoretical relation $\nu(\tau) = \frac{1}{4} \left( \tau - \frac{\delta t}{2} \right)$ shows good 
agreement with ultra-reativistic models based on exact streaming~\cite{furtmaier15} but the interpolated streaming introduces a numerical
diffusivity which increases the effective viscosity of the fluid~\cite{PhysRevE.61.6546, Wu20112246, Yu2003329}, i.e.,
\begin{eqnarray}
\nu_{eff} = \frac{1}{4}\left[ \tau  - \delta t\left(\frac{1}{2} + \delta_\nu \right) \right] .
\end{eqnarray}
By linear fit, we measure $\delta_\nu = - 0.2454 \pm 0.0002$. This relation is in good agreement with the data from simulations as can be seen in Fig. \ref{eta-tau-fig}. One can find the shear viscosity by $\eta=\nu (\varepsilon + P)$. For realistic simulations, the relaxation time should not be constant. Instead the shear viscosity to entropy ratio ($\eta/s$) should be constant~\cite{muller09} which is accounted in the simulations for the KHI in graphene and therefore the relaxation time reads $\tau = 4 \eta/(s \theta) + 0.2546\,\delta t $ for a relative temperature $\theta = T/T_0$.

\subsection{Thermal conductivity measurement}

\begin{figure}[htb]
\center
\includegraphics[width= \linewidth]{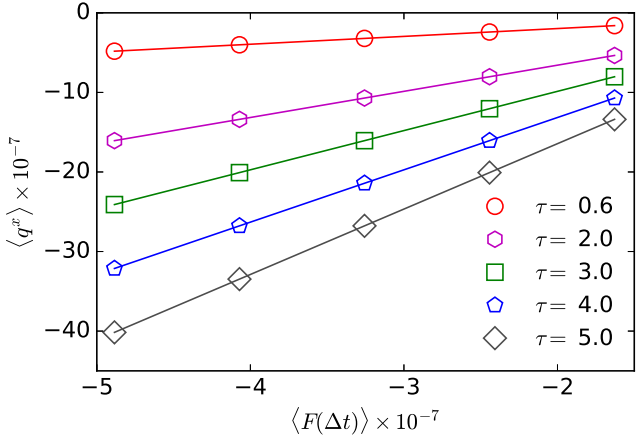}
\caption{(Color online) Average heat flux as a function of $\langle F(\Delta T) \rangle $ (function of the temperature gradient)
for 5 different relaxation times. The solid lines are linear fits for each relaxation time.}
\label{qx-dt}
\end{figure}

\begin{figure}[htb]
\center
\includegraphics[width= \linewidth]{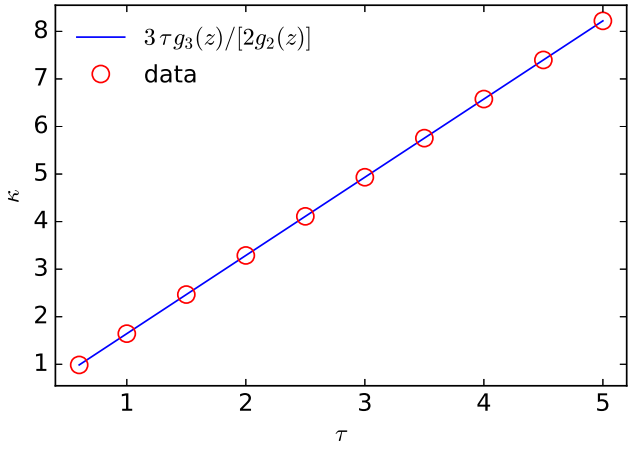}
\caption{(Color online) Thermal conductivity for 10 different relaxation times compared with the function in Eq.\eqref{kappa-tau-eq}. }
\label{kappa-tau}
\end{figure}

The heat flux can be related to the thermal conductivity by~\cite{cercignani02}
\begin{eqnarray}\label{heat-flux-eq}
q^\alpha = \kappa \left(\nabla^\alpha T - \frac{T}{v_F^2}D U^\alpha \right)
\end{eqnarray}
where $D= U^\alpha \partial_\alpha$. To measure the thermal conductivity we simulate an effectively one dimensional system of size $L_x\times L_y = 2048\times 2$ with open boundary conditions (except by the temperature, which is set constant) on left and right and periodic boundary on top and bottom for 5 different gradients in temperature in the $x$ direction and we calculate the heat flux (see Eq.\eqref{charge-flow-eq}),
\begin{eqnarray}
q^\alpha = \frac{3T g_3(z)}{g_2(z)}( n U^\alpha - N^\alpha).
\end{eqnarray}
For a one dimensional gradient, Eq.\eqref{heat-flux-eq} becomes
\begin{eqnarray}
q^x =\kappa F(\Delta T) 
\end{eqnarray}
where
\begin{eqnarray}
F(\Delta T) \equiv &&-  \left\{ \left( 1+\frac{(u^x)^2 \gamma^2}{v_F^2} \right) \frac{\partial T}{\partial x} \right.\\ \nonumber&&
\left. +\frac{T\gamma}{v_F^2}
\left[ v_F\frac{\partial}{\partial t}(\gamma u^x) + u^x \frac{\partial}{\partial x}(\gamma u^x)  \right]      \right\}.
\end{eqnarray}
In the classical limit, Eq.\eqref{heat-flux-eq} becomes Fourier's law, while $F(\Delta T) \rightarrow - \partial T/\partial x$. We calculate the spatial average of $F$, $\langle  F(\Delta T) \rangle$, and the average heat flux, $\langle q^x   \rangle$ (both are essentially
constant in space) for 5 different temperature gradients ($\Delta T$). For each simulation, the temperature on the boundaries is set as $\theta_L = 1-\Delta T/2$ on the left and $\theta_R=1+\Delta T /2$ on the right, while the differences in temperature are
$\Delta T = \{5.0,\, 7.5,\, 10.0, \,12.5,\, 15.0\}\times 10^{-4}$. The initial conditions are $n_0=1.0$ and $\mathbf{u_0}=0$
everywhere and we set an initial temperature gradient as $\theta_0(x) =  \theta_L + x (\theta_R - \theta_L)/L_x$. 
Fig. \ref{qx-dt} shows the average heat flux as a function of $\langle F(\Delta T) \rangle $ for 5 relaxation times  and its respective linear fits. The slope of each line gives the thermal conductivity,
which can be seen in Fig. \ref{kappa-tau} as a function of the relaxation time. The linear fits suggest that the relation 
between the thermal conductivity and the relaxation time is 
\begin{eqnarray}\label{kappa-tau-eq}
\kappa(\tau) = \frac{3\,\tau \,g_3(z)}{2\, g_2(z)},
\end{eqnarray}
which is close to the relation found in Ref. ~\cite{furtmaier15}, but with better accuracy since we are using a fifth order expansion.


\section{Kelvin-Helmholtz instability}\label{khi-sec}

When two fluid or two regions of the same fluid shear against each other with different tangential velocities and a perturbation is introduced on the interface, the KHI takes place. To understand the critical values for which the instability occurs, lets consider two fluids, separated by a flat interface in the middle, under an external force
perpendicular to the velocities, e.g, an electrical force~\cite{furtmaier15}. The fluid in the upper part has smaller energy density $\varepsilon_2$ and is moving with velocity $\mathbf{U_2}$ while the fluid in the bottom has energy density $\varepsilon_1$ and
velocity $\mathbf{U_1}$. If a perturbation in the fields (charge density, velocity or pressure), 
\begin{eqnarray}
\delta q \propto \exp [i(k x + l y - \omega t)],
\end{eqnarray}
is introduced at the interface, a linear stability
analysis~\cite{chandrasekhar61} provides that the minimum  
wave number in the parallel direction (transverse modes do not affect the instability) of the shear flow to have the KHI is 
\begin{eqnarray}\label{kmin}
k_{min} = \frac{E\, g_2(z)|\varepsilon _1^2 - \varepsilon _2^2|}{3\,\varepsilon _1 \,\varepsilon _2\,T\,g_3(z)(U_1-U_2)^2},
\end{eqnarray}
where we considered an external electrical field $E$ perpendicular to the flow causing an acceleration $\frac{n E }{\varepsilon+P} = \frac{E g_2(z)}{3Tg_3(z)}$. The KHI occurs for any $k > k_{min}$. Note that here the external force has a stabilizing role. Another way to stabilize the shear flow is with a gradient of
charge density and/or velocity~\cite{gan11}. Defining the relativistic Richardson number for this problem as 
\begin{eqnarray}
Ri = - \frac{E\,g_2(z)}{3\,\varepsilon\, T\,g_3(z)} \frac{d \varepsilon /dy}{(dU^x/dy)^2},
\end{eqnarray}
the linear stability analysis gives that the necessary condition to have a stable flow is $Ri > 1/4$ everywhere~\cite{nakayama90, chandrasekhar61}. The flow can be stable for $Ri<1/4$ only in the
absence of perturbations. The flow can also be stable for supersonic shear velocities~\cite{bodo04}. For instance, for the simple case with $l=\omega=0$, the flow is stable when $\mathcal{M}>1$, where the relativistic Mach number is defined as
\begin{eqnarray}
\mathcal{M} = \frac{u^x \gamma(u^x)}{c_s \gamma(c_s)}.
\end{eqnarray}
For the conditions we consider in the simulations for graphene, the flow is unstable for every perturbation because we do not have any external force perpendicular to the flow neither supersonic velocities.

\begin{figure}[ht]
\center
\includegraphics[width=\linewidth]{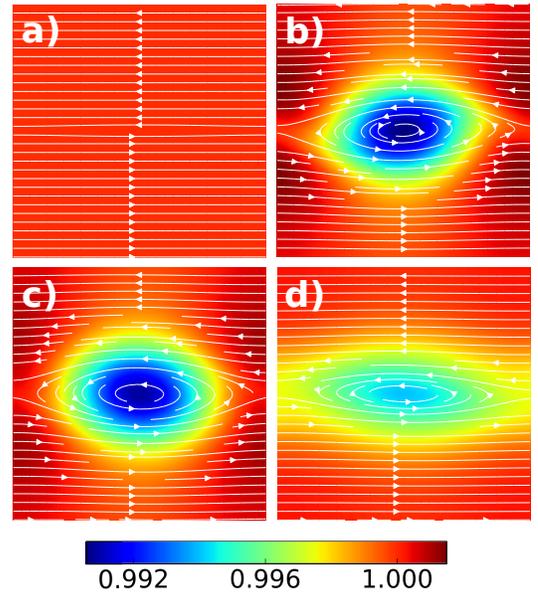}
\caption{(Color online) Formation of the KHI in graphene at a) $t =0$ ns, b) $t=0.72$ ns, c) $t=1.00$ ns, d) $t=1.43$ ns. The streamlines shows the velocity field and the colors represents the charge density fluctuations relative to the initial charge density, $n/n_0$.}
\label{cat-eye}
\end{figure}

In the following simulations, we consider that the charge carriers are in the hydrodynamic regime, which implies that the mean free path for carrier-carrier collisions gives the smallest spatial scale for the system. See  Ref.~\cite{kumar2017super} for measurements of mean free paths and for the transition between ballistic and hydrodynamic regime in graphene. In order to reduce the scattering with impurities and phonon, we consider ultra-clean samples at appropriate temperature. The sample is on a substrate, e.g., SiO$_2$,  with finite carrier density controlled by an external gate voltage. In addition, all simulations are performed for small velocities.

\subsection{Ideal setup}

As an idealized setup to observe the KHI, we model a system with size $L_x \times L_y = 512 \times 512$ grid points, representing a $37\mu \mbox{m} \times 37\mu \mbox{m}$ physical system, where the fluid has opposite velocities in the two halves, that is, 
\begin{eqnarray}
u_x^0 = -U_0\tanh\left(\frac{y-L_y/2}{a} \right),
\end{eqnarray}
where we set $U_0 = 0.1 v_F$ and $a = 1$. We introduce a small perturbation to trigger the instability as 
\begin{eqnarray}
u_y^0 = u_{pert} \sin\left[\frac{2\pi (x-L_x/2)}{L_x}\right] \exp\left[-\frac{(y-L_y/2)^2}{b^2}\right],\nonumber \\
\end{eqnarray}
where $u_{pert} = 0.005 v_F$ and $b = 10$. Initially, the charge density~\cite{mendoza13} and the temperature are the same everywhere, $n_0=2.26\times10^{-5}$ C/m$^{2}$ and $T_0 = 100$ K. For this temperature, the electron-phonon interactions are negligible~\cite{barreiro09}. The
numerical shear viscosity-entropy ratio for the simulations of the KHI is $\eta/s = 0.12$. By using the Gibbs-Duhem relation for zero chemical potential, $\varepsilon + p = s T$, we calculate the kinematic viscosity $\nu = (\eta/s)/T_0=0.12$ and the Reynolds number for this simulation, $Re = L_0 v_0/\nu = 427$, where we use the size of the sample as the characteristic length $L_0$ and the velocity in each half as the characteristic velocity $v_0$. For a graphene sample with $T=100 K$ the kinematic viscosity~\cite{mendoza11} is $\nu = 8.57 \times 10^{-3}\, m^2/s$. The boundary conditions are periodic in left and right direction and, at top and bottom, the boundary is open except for the horizontal velocity $u_x (t) = u_x^0$ that is set constant. In Fig. \ref{cat-eye} we see the formation and evolution of the KHI for different times ($\delta t = 71$ fs). At $t=0$ ns, we have the two regions of the fluid moving in opposite directions and a small perturbation in the velocity field at the middle. Since there is no external force perpendicular to the flow, Eq.\eqref{kmin} gives that $k_{min}=0$, i.e., any perturbation makes the flow be unstable. Therefore the KHI appears as we can see in Fig. \ref{cat-eye} for $t=0.72$ ns and $t=1.00$ ns, where we can recognize the pattern of the cat-eyes in the charge density field. After some time, the flow stabilizes due to the generation of a gradient in the velocity and charge density fields and to the absence of perturbations (Fig.\ref{cat-eye} d). 

\subsection{Realistic setup}
\begin{figure}[ht]
\center
\includegraphics[width=\linewidth]{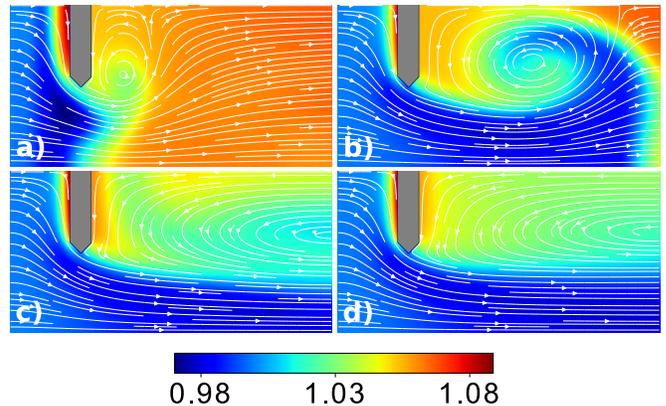}
\caption{(Color online) Realistic setup to observe the KHI at $Re=53$. By using a constant current $0.05 \, v_F$ in the source (left side), we see the snapshots for a) $t=0.14$ ns, b) $t=0.43$ ns, c) $t=0.85$ ns, d) $t=1.42$ ns. The colors represent the density fluctuations relative to the initial density, $n/n_0$, and the gray object represents a needle shaped obstacle. The streamlines show the directions of the velocity field.}
\label{obst}
\end{figure}
\begin{figure}[ht]
\center
\includegraphics[width=\linewidth]{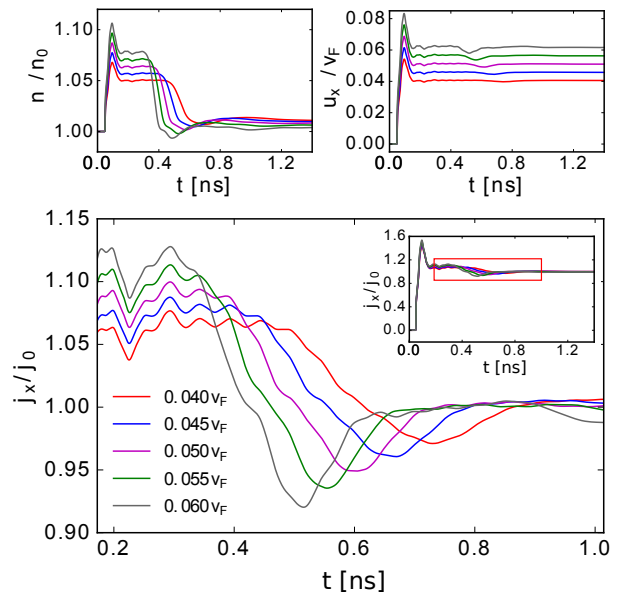}
\caption{(Color online) Average charge density, average x-component of the velocity and current as functions of time for different source velocities measured close to the drain. The inset show the current for a longer time and the red rectangle indicates the region that is being amplified.}
\label{fields}
\end{figure}

\begin{figure}[ht]
\center
\includegraphics[width=\linewidth]{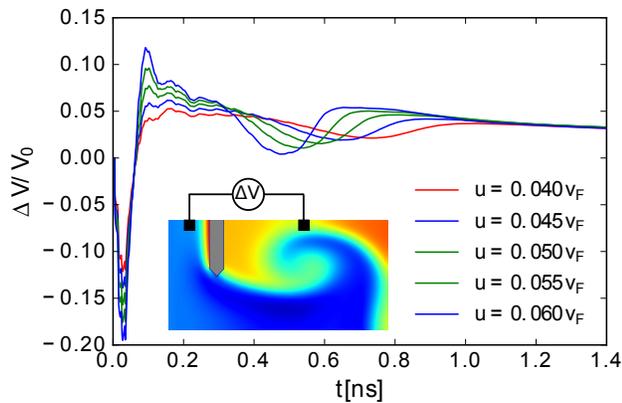}
\caption{(Color online) EPD between the two points indicated by the black squares in the inset divided by a reference voltage $V_0$.}
\label{ddp-inset}
\end{figure}
In order to detect the KHI in experiments we propose a more realistic setup that could be performed nowadays, where we force the Dirac fluid to flow through an obstacle (see Fig. \ref{obst}). We simulate a system 
with $L_x \times L_y = 512 \times 256$ with a needle shaped obstacle measuring $16\times 128$ nodes, which represents $1.1 \,\mu \mbox{m} \times 9.1 \,\mu \mbox{m}$, positioned 96 nodes ($6.8\,\mu $m) away from the source. Initially, all fields are homogeneous: $n_0=2.26\times10^{-5}$ C/m$^{2}$, $T_0=100$ K, $\mathbf{u_0} = 0$. We
use bounce-back boundary conditions at the obstacle's surface ($\mathbf{u}=0$), open boundary at the right side (drain), slide-free boundaries at top and bottom ($u_y=0$) and, at the left side, the source, we set a current in the horizontal direction: $n^{in} = n_0$, $u_x^{in}(t)$, $u_y^{in} = 0$, and we obtain the temperature at the boundaries by a zero-order extrapolation from the first fluid neighbors. 
\begin{figure}[ht]
\center
\includegraphics[width=\linewidth]{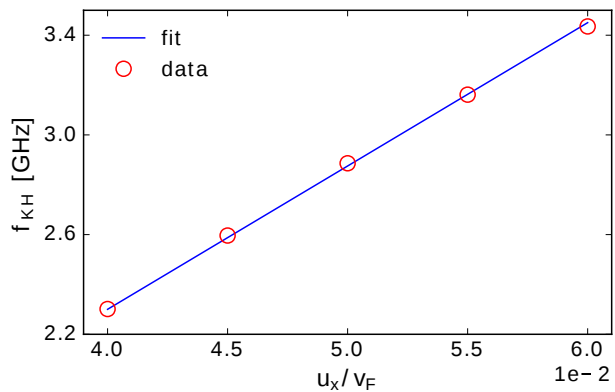}
\caption{(Color online) Frequency of the signal due to the KHI as a function of the source velocity and a linear fit.}
\label{fkh}
\end{figure}
Now we analyze the fields when a constant current is applied at the source. In Fig. \ref{obst}, we see the evolution of the charge density field and the formation of the KHI for a velocity $u_x^{in} = 0.05v_F$ at the source, which corresponds to an electrical current of $I_0 = j_0 \,L_y = 20.6\,\mu$A. Considering $u_x^{in}$ as the characteristic velocity and the length of the obstacle as the characteristic
length, we have $Re = 53$ for this simulation. When the current reaches the obstacle ($t=0.14$ ns), we see that the fluid at the bottom region has velocity $>u_x^{in}$, while the fluid at the upper region has $u_x = 0$ generating a shear flow. Since we have no external force in the vertical direction, Eq.\eqref{kmin} says that the flow is
unstable for every perturbation, which, in our case, is generated by the initial passage of the fluid and, therefore, the KHI appears (Fig. \ref{obst}b). At $t=0.85$ ns the flow begins to stabilize due to the formation of gradients and the absence of perturbations and, at $t=1.42$ ns, we can not see signs of the instability anymore. The streamlines in Fig. \ref{obst} show that, after the passage of the KHI, we have the formation of permanent (steady state) whirlpool-like regions between the obstacle and the drain similarly to the ones reported in Refs.~\cite{PhysRevB.92.165433, pellegrino2016electron, levitov16, bandurin16}. It suggests that the KHI drives the formation of these experimentally observed whirlpools in graphene analogously to many other vortex formation in nature~\cite{smyth2012ocean, PhysRevLett.62.772, wyper2013kelvin}. The KHI can be identified in the electrical current signal, because there are
fluctuations in charge density and velocity when the instability passes by the measurement points. In Fig. \ref{fields}, we see the time evolution for the current , $j_x(t) = \int dy\, n\,u_x$, the average charge density, $ n(t)  = \int dy \, n /L_y$, and the average x-component of the velocity, 
$u_x(t)  = \int dy\,u_x/L_y$, measured close to the drain (10 nodes before) for 5 source velocities, where $\delta t = 71$ fs. For the velocity $0.05v_F$, we can observe
fluctuations in the fields due to the instability starting approximately from 0.36 ns to 0.71 ns, which agrees with, respectively, the times when the instability reaches the right border and disappears in Fig. \ref{obst}. In the inset of Fig. \ref{fields} one can observe the first big oscillation in the electrical
current that is due to the waves generated by the initial passage of the fluid through the obstacle. Since these waves depend only of the sound speed, they reach the drain at the same time, independently of the source velocity $u_x^{in}$. After this, one can observe oscillations, of few microamperes, due to the KHI that have a smaller period for higher source velocities. This is expected as the instabilities have approximately the same dimensions, but travel faster for higher velocities. To estimate the period of each oscillation of the instability,
$T_{KH}$, we consider the charge density curves, since they are smoother and the instability's sign can be identified more easily. In order to numerically measure the beginning of the oscillation, we define it as the point at which the derivative is smaller than a
reference value, which we choose as being half of the derivative at the decreasing region in the fields (for instance, between 0.4 ns and 0.6 ns for $u_0 = 0.05 v_F$). We find the end of the oscillation in a analogous way but considering the derivative in the increasing region. Thus, we calculate the frequency of the instability defined by $f_{KH}= 1/T_{KH}$ and plot it as a function of the source velocity, Fig. \ref{fkh}. We can identify a linear relation, which is expected from the wave equation $v=\lambda \times f$. By a linear fit we find $\lambda \approx 17.4 \,\mu $m, that approximately corresponds to the length of the
instability. In Fig. \ref{obst}, we see that the length of the instability does correspond to roughly half of the system size ($18.2\mu m$), what confirms that this oscillation in the current measurement is due to the KHI. 

One can detect the instability in experiments by measuring the electric potential difference (EPD). We consider the simplification adopted in Ref.~\cite{torre15}, which considers that the EPD is caused by fluctuations in the charge density field, leading to:
\begin{eqnarray}
 \Delta V \approx  \frac{\Delta n}{C},
\end{eqnarray}
where $C = \epsilon _0 \epsilon _r / d$ is the capacitance per unit area, $\epsilon _0$ is the vacuum permeability, $\epsilon _r$ is the relative permeability of the substrate and $d$ is the thickness of the substrate. Fig. \ref{ddp-inset} shows the EPD between the two points indicated by black squares in the inset (upper boundary and in the middle of each domain) divided by a reference potential, $V_0= n_0 / C$, with $n_0$ being the initial density. Here, $\Delta n = n_R-n_L$ is the difference between the charge density at the right and left contacts. Initially, the EPD is zero, due to the homogeneous initial condition in charge density. The first oscillations occur when the moving fluid reaches the contacts and they do not depend on the fluid velocity as discussed before. Between 0.3 ns and 1 ns, we can see the oscilations due to the KHI, which depend on the fluid velocity likewise with the electrical current. Considering, for instance, a substrate of SiO$_2$, which has $\epsilon_r=3.9$, and typical experimental parameters~\cite{dorgan10} ($d = 3\times 10^{-7}$ m, $n_0=2.26\times10^{-5}$ C/m$^{2}$), we can estimate that the oscillations due to the KHI are on the scale of $\sim 10$ mV, which could be measured in current experiments. The oscillations in the electrical current, on the scale of microamperes, are much harder to detect.  

\begin{figure}[ht]
\center
\includegraphics[width=\linewidth]{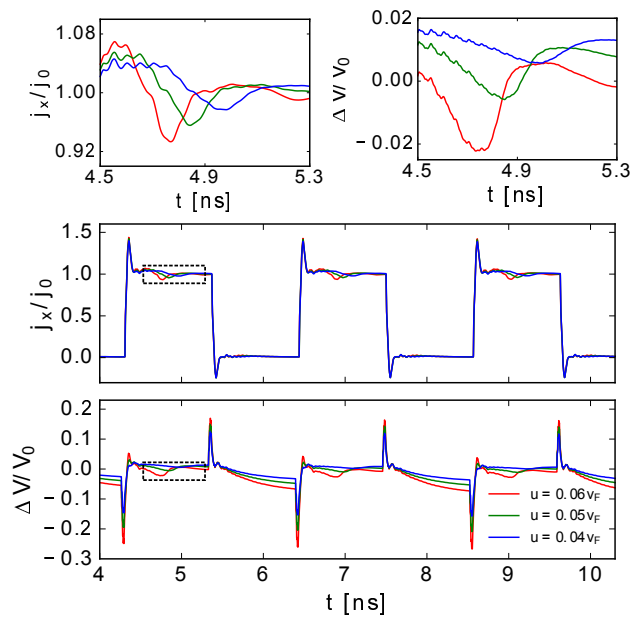}
\caption{(Color online) Electrical current at the drain and EPD between the two contacts indicated in Fig. \ref{ddp-inset} when an alternated squared current of 470 MHz is applied at the source for three different source velocities. The dashed rectangles indicate where the oscillations due to the KHI can be identified, which are amplified in the figures on the top. }
\label{alternate}
\end{figure}

Since the duration of the KHI is on the scale of nanoseconds, it would be challenging to observe it with a constant current, but one could generate it with a high frequency and observe its influence on the electrical current and EPD. We
simulate a squared current (on-off) with a frequency of $470$ MHz for three source velocities and the time dependence of the electrical current and the EPD can be seen in Fig. \ref{alternate} for three cycles starting from 4 ns to avoid the initial stabilization of the system. The behavior that we observed for a constant current (Figs. \ref{fields} and \ref{ddp-inset}) can be reproduced indefinitely and we can clearly identify the oscillations that are due to the KHI,
since they change with the source velocity. As can be seen in Fig. \ref{alternate}, the cycles are basically identical and, therefore, one could distinguish the oscillations due to the KHI from the experimental noise by taking the statistical average of many cycles. Note that the current at the drain becomes negative when the source current is interrupted, which is due to the whirlpools (see Fig. \ref{obst}) that cause a back flow.

\section{Conclusions}\label{conclusions}

The Kelvin-Helmholtz instability was analyzed in an idealized setup, with a shear flow between two regions of the Dirac fluid moving in opposite directions. We also simulated a flow through a needle shaped obstacle, which would be a possible experimental realization to observe this instability, and we analyzed its impact on the electrical potential difference measurements. The Kelvin-Helmholtz instability can be identified by changing the current at the source. An alternating squared current can be used to produce the instability many times, such that one can later take the statistical average over the different cycles and differentiate the instability from noise. Since this instability always occur in the presence of an obstacle, it can even be produced and measured accidentally in experiments and be confused with experimental noise. Therefore, it should be considered in experiments performing measurements on the scale of nanoseconds. As illustrated here, the Kelvin-Helmholtz instability leads to the formation of whirlpools similar to the ones reported in Ref.~\cite{bandurin16} (see Fig. \ref{obst}). 

A new lattice Boltzmann method based on the fifth order expansion of the Fermi-Dirac distribution was proposed and applied to study the Kelvin-Helmholtz instability on graphene. The expansion was made in relativistic polynomials specifically developed to expand the relativistic Fermi-Dirac distribution in two dimensions, but the method described here could be straightforwardly generalized to other distribution functions, as the Maxwell-J\"uttner distribution, and also to three dimensions, since the polynomials are written in a general tensorial form. Also a new quadrature that is able to calculate up to the fifth order moment was developed for this model. This quadrature has the disadvantage to use an interpolation in the streaming step, but it keeps a high grid resolution, which is a problem for the previous model with improved dissipation for a third order expansion~\cite{mendoza13-3}.  The fifth order expansion provides the full set of conservation equations for a fluid, which is necessary to describe accurately viscous effects as the Kelvin-Helmholtz instability. The model was validated by the Riemann problem and characterized in order to find the relation between the viscosity and thermal conductivity with the relaxation time. 

Although we have considered the Dirac fluid in graphene, the analysis and model presented in this work could be extended to a broader class of the Dirac materials~\cite{wehling14}. It opens the way to investigate hydrodynamic effects on these novel materials, including topological insulators~\cite{PhysRevB.93.155122}, which has carriers on the surface that may behave like a fluid, Weyl systems~\cite{Lucas23082016} and 2D metal Palladium cobaltate~\cite{Moll1061}.

\begin{acknowledgements}
R.C.V. Coelho, M. Mendoza and H. J. Herrmann thank to the European Research Council (ERC) Advanced Grant 319968-FlowCCS and R.C.V. Coelho thanks to FAPERJ for the financial support. The authors are thankful with Prof. Klaus Ensslin and his team for fruitful discussions about experimental realizations.
\end{acknowledgements}

\bibliography{references}

\begin{thebibliography}{75}%
\makeatletter
\providecommand \@ifxundefined [1]{%
 \@ifx{#1\undefined}
}%
\providecommand \@ifnum [1]{%
 \ifnum #1\expandafter \@firstoftwo
 \else \expandafter \@secondoftwo
 \fi
}%
\providecommand \@ifx [1]{%
 \ifx #1\expandafter \@firstoftwo
 \else \expandafter \@secondoftwo
 \fi
}%
\providecommand \natexlab [1]{#1}%
\providecommand \enquote  [1]{``#1''}%
\providecommand \bibnamefont  [1]{#1}%
\providecommand \bibfnamefont [1]{#1}%
\providecommand \citenamefont [1]{#1}%
\providecommand \href@noop [0]{\@secondoftwo}%
\providecommand \href [0]{\begingroup \@sanitize@url \@href}%
\providecommand \@href[1]{\@@startlink{#1}\@@href}%
\providecommand \@@href[1]{\endgroup#1\@@endlink}%
\providecommand \@sanitize@url [0]{\catcode `\\12\catcode `\$12\catcode
  `\&12\catcode `\#12\catcode `\^12\catcode `\_12\catcode `\%12\relax}%
\providecommand \@@startlink[1]{}%
\providecommand \@@endlink[0]{}%
\providecommand \url  [0]{\begingroup\@sanitize@url \@url }%
\providecommand \@url [1]{\endgroup\@href {#1}{\urlprefix }}%
\providecommand \urlprefix  [0]{URL }%
\providecommand \Eprint [0]{\href }%
\providecommand \doibase [0]{http://dx.doi.org/}%
\providecommand \selectlanguage [0]{\@gobble}%
\providecommand \bibinfo  [0]{\@secondoftwo}%
\providecommand \bibfield  [0]{\@secondoftwo}%
\providecommand \translation [1]{[#1]}%
\providecommand \BibitemOpen [0]{}%
\providecommand \bibitemStop [0]{}%
\providecommand \bibitemNoStop [0]{.\EOS\space}%
\providecommand \EOS [0]{\spacefactor3000\relax}%
\providecommand \BibitemShut  [1]{\csname bibitem#1\endcsname}%
\let\auto@bib@innerbib\@empty
\bibitem [{\citenamefont {Novoselov}\ \emph {et~al.}(2004)\citenamefont
  {Novoselov}, \citenamefont {Geim}, \citenamefont {Morozov}, \citenamefont
  {Jiang}, \citenamefont {Zhang}, \citenamefont {Dubonos}, \citenamefont
  {Grigorieva},\ and\ \citenamefont {Firsov}}]{novoselov04}%
  \BibitemOpen
  \bibfield  {author} {\bibinfo {author} {\bibfnamefont {K.~S.}\ \bibnamefont
  {Novoselov}}, \bibinfo {author} {\bibfnamefont {A.~K.}\ \bibnamefont {Geim}},
  \bibinfo {author} {\bibfnamefont {S.~V.}\ \bibnamefont {Morozov}}, \bibinfo
  {author} {\bibfnamefont {D.}~\bibnamefont {Jiang}}, \bibinfo {author}
  {\bibfnamefont {Y.}~\bibnamefont {Zhang}}, \bibinfo {author} {\bibfnamefont
  {S.~V.}\ \bibnamefont {Dubonos}}, \bibinfo {author} {\bibfnamefont {I.~V.}\
  \bibnamefont {Grigorieva}}, \ and\ \bibinfo {author} {\bibfnamefont {A.~A.}\
  \bibnamefont {Firsov}},\ }\href {\doibase 10.1126/science.1102896} {\bibfield
   {journal} {\bibinfo  {journal} {Science}\ }\textbf {\bibinfo {volume}
  {306}},\ \bibinfo {pages} {666} (\bibinfo {year} {2004})}\BibitemShut
  {NoStop}%
\bibitem [{\citenamefont {Novoselov}\ \emph {et~al.}(2005)\citenamefont
  {Novoselov}, \citenamefont {Geim}, \citenamefont {Morozov}, \citenamefont
  {Jiang}, \citenamefont {Katsnelson}, \citenamefont {Grigorieva},
  \citenamefont {Dubonos},\ and\ \citenamefont {Firsov}}]{novoselov05}%
  \BibitemOpen
  \bibfield  {author} {\bibinfo {author} {\bibfnamefont {K.~S.}\ \bibnamefont
  {Novoselov}}, \bibinfo {author} {\bibfnamefont {A.~K.}\ \bibnamefont {Geim}},
  \bibinfo {author} {\bibfnamefont {S.~V.}\ \bibnamefont {Morozov}}, \bibinfo
  {author} {\bibfnamefont {D.}~\bibnamefont {Jiang}}, \bibinfo {author}
  {\bibfnamefont {M.~I.}\ \bibnamefont {Katsnelson}}, \bibinfo {author}
  {\bibfnamefont {I.~V.}\ \bibnamefont {Grigorieva}}, \bibinfo {author}
  {\bibfnamefont {S.~V.}\ \bibnamefont {Dubonos}}, \ and\ \bibinfo {author}
  {\bibfnamefont {A.~A.}\ \bibnamefont {Firsov}},\ }\href
  {http://dx.doi.org/10.1038/nature04233} {\bibfield  {journal} {\bibinfo
  {journal} {Nature}\ }\textbf {\bibinfo {volume} {438}},\ \bibinfo {pages}
  {197} (\bibinfo {year} {2005})}\BibitemShut {NoStop}%
\bibitem [{\citenamefont {Castro~Neto}\ \emph {et~al.}(2009)\citenamefont
  {Castro~Neto}, \citenamefont {Guinea}, \citenamefont {Peres}, \citenamefont
  {Novoselov},\ and\ \citenamefont {Geim}}]{RevModPhys.81.109}%
  \BibitemOpen
  \bibfield  {author} {\bibinfo {author} {\bibfnamefont {A.~H.}\ \bibnamefont
  {Castro~Neto}}, \bibinfo {author} {\bibfnamefont {F.}~\bibnamefont {Guinea}},
  \bibinfo {author} {\bibfnamefont {N.~M.~R.}\ \bibnamefont {Peres}}, \bibinfo
  {author} {\bibfnamefont {K.~S.}\ \bibnamefont {Novoselov}}, \ and\ \bibinfo
  {author} {\bibfnamefont {A.~K.}\ \bibnamefont {Geim}},\ }\href {\doibase
  10.1103/RevModPhys.81.109} {\bibfield  {journal} {\bibinfo  {journal} {Rev.
  Mod. Phys.}\ }\textbf {\bibinfo {volume} {81}},\ \bibinfo {pages} {109}
  (\bibinfo {year} {2009})}\BibitemShut {NoStop}%
\bibitem [{\citenamefont {M\"uller}\ \emph {et~al.}(2009)\citenamefont
  {M\"uller}, \citenamefont {Schmalian},\ and\ \citenamefont
  {Fritz}}]{muller09}%
  \BibitemOpen
  \bibfield  {author} {\bibinfo {author} {\bibfnamefont {M.}~\bibnamefont
  {M\"uller}}, \bibinfo {author} {\bibfnamefont {J.}~\bibnamefont {Schmalian}},
  \ and\ \bibinfo {author} {\bibfnamefont {L.}~\bibnamefont {Fritz}},\ }\href
  {\doibase 10.1103/PhysRevLett.103.025301} {\bibfield  {journal} {\bibinfo
  {journal} {Phys. Rev. Lett.}\ }\textbf {\bibinfo {volume} {103}},\ \bibinfo
  {pages} {025301} (\bibinfo {year} {2009})}\BibitemShut {NoStop}%
\bibitem [{\citenamefont {Balandin}\ \emph {et~al.}(2008)\citenamefont
  {Balandin}, \citenamefont {Ghosh}, \citenamefont {Bao}, \citenamefont
  {Calizo}, \citenamefont {Teweldebrhan}, \citenamefont {Miao},\ and\
  \citenamefont {Lau}}]{baladin08}%
  \BibitemOpen
  \bibfield  {author} {\bibinfo {author} {\bibfnamefont {A.~A.}\ \bibnamefont
  {Balandin}}, \bibinfo {author} {\bibfnamefont {S.}~\bibnamefont {Ghosh}},
  \bibinfo {author} {\bibfnamefont {W.}~\bibnamefont {Bao}}, \bibinfo {author}
  {\bibfnamefont {I.}~\bibnamefont {Calizo}}, \bibinfo {author} {\bibfnamefont
  {D.}~\bibnamefont {Teweldebrhan}}, \bibinfo {author} {\bibfnamefont
  {F.}~\bibnamefont {Miao}}, \ and\ \bibinfo {author} {\bibfnamefont {C.~N.}\
  \bibnamefont {Lau}},\ }\href {\doibase 10.1021/nl0731872} {\bibfield
  {journal} {\bibinfo  {journal} {Nano Letters}\ }\textbf {\bibinfo {volume}
  {8}},\ \bibinfo {pages} {902} (\bibinfo {year} {2008})}\BibitemShut {NoStop}%
\bibitem [{\citenamefont {Dorgan}\ \emph {et~al.}(2010)\citenamefont {Dorgan},
  \citenamefont {Bae},\ and\ \citenamefont {Pop}}]{dorgan10}%
  \BibitemOpen
  \bibfield  {author} {\bibinfo {author} {\bibfnamefont {V.~E.}\ \bibnamefont
  {Dorgan}}, \bibinfo {author} {\bibfnamefont {M.-H.}\ \bibnamefont {Bae}}, \
  and\ \bibinfo {author} {\bibfnamefont {E.}~\bibnamefont {Pop}},\ }\href
  {\doibase 10.1063/1.3483130} {\bibfield  {journal} {\bibinfo  {journal}
  {Applied Physics Letters}\ }\textbf {\bibinfo {volume} {97}},\ \bibinfo
  {pages} {082112} (\bibinfo {year} {2010})}\BibitemShut {NoStop}%
\bibitem [{\citenamefont {de~Jong}\ and\ \citenamefont
  {Molenkamp}(1995)}]{jong1995hydrodynamic}%
  \BibitemOpen
  \bibfield  {author} {\bibinfo {author} {\bibfnamefont {M.~J.~M.}\
  \bibnamefont {de~Jong}}\ and\ \bibinfo {author} {\bibfnamefont {L.~W.}\
  \bibnamefont {Molenkamp}},\ }\href {\doibase 10.1103/PhysRevB.51.13389}
  {\bibfield  {journal} {\bibinfo  {journal} {Phys. Rev. B}\ }\textbf {\bibinfo
  {volume} {51}},\ \bibinfo {pages} {13389} (\bibinfo {year}
  {1995})}\BibitemShut {NoStop}%
\bibitem [{\citenamefont {Gurzhi}(1968)}]{gurzhi1968hydrodynamic}%
  \BibitemOpen
  \bibfield  {author} {\bibinfo {author} {\bibfnamefont {R.~N.}\ \bibnamefont
  {Gurzhi}},\ }\href {http://stacks.iop.org/0038-5670/11/i=2/a=R07} {\bibfield
  {journal} {\bibinfo  {journal} {Sov. Phys. Usp.}\ }\textbf {\bibinfo {volume}
  {11}},\ \bibinfo {pages} {255} (\bibinfo {year} {1968})}\BibitemShut
  {NoStop}%
\bibitem [{\citenamefont {Bandurin}\ \emph {et~al.}(2016)\citenamefont
  {Bandurin}, \citenamefont {Torre}, \citenamefont {Kumar}, \citenamefont
  {Ben~Shalom}, \citenamefont {Tomadin}, \citenamefont {Principi},
  \citenamefont {Auton}, \citenamefont {Khestanova}, \citenamefont {Novoselov},
  \citenamefont {Grigorieva}, \citenamefont {Ponomarenko}, \citenamefont
  {Geim},\ and\ \citenamefont {Polini}}]{bandurin16}%
  \BibitemOpen
  \bibfield  {author} {\bibinfo {author} {\bibfnamefont {D.~A.}\ \bibnamefont
  {Bandurin}}, \bibinfo {author} {\bibfnamefont {I.}~\bibnamefont {Torre}},
  \bibinfo {author} {\bibfnamefont {R.~K.}\ \bibnamefont {Kumar}}, \bibinfo
  {author} {\bibfnamefont {M.}~\bibnamefont {Ben~Shalom}}, \bibinfo {author}
  {\bibfnamefont {A.}~\bibnamefont {Tomadin}}, \bibinfo {author} {\bibfnamefont
  {A.}~\bibnamefont {Principi}}, \bibinfo {author} {\bibfnamefont {G.~H.}\
  \bibnamefont {Auton}}, \bibinfo {author} {\bibfnamefont {E.}~\bibnamefont
  {Khestanova}}, \bibinfo {author} {\bibfnamefont {K.~S.}\ \bibnamefont
  {Novoselov}}, \bibinfo {author} {\bibfnamefont {I.~V.}\ \bibnamefont
  {Grigorieva}}, \bibinfo {author} {\bibfnamefont {L.~A.}\ \bibnamefont
  {Ponomarenko}}, \bibinfo {author} {\bibfnamefont {A.~K.}\ \bibnamefont
  {Geim}}, \ and\ \bibinfo {author} {\bibfnamefont {M.}~\bibnamefont
  {Polini}},\ }\href {\doibase 10.1126/science.aad0201} {\bibfield  {journal}
  {\bibinfo  {journal} {Science}\ }\textbf {\bibinfo {volume} {351}},\ \bibinfo
  {pages} {1055} (\bibinfo {year} {2016})}\BibitemShut {NoStop}%
\bibitem [{\citenamefont {Tikhonenko}\ \emph {et~al.}(2009)\citenamefont
  {Tikhonenko}, \citenamefont {Kozikov}, \citenamefont {Savchenko},\ and\
  \citenamefont {Gorbachev}}]{PhysRevLett.103.226801}%
  \BibitemOpen
  \bibfield  {author} {\bibinfo {author} {\bibfnamefont {F.~V.}\ \bibnamefont
  {Tikhonenko}}, \bibinfo {author} {\bibfnamefont {A.~A.}\ \bibnamefont
  {Kozikov}}, \bibinfo {author} {\bibfnamefont {A.~K.}\ \bibnamefont
  {Savchenko}}, \ and\ \bibinfo {author} {\bibfnamefont {R.~V.}\ \bibnamefont
  {Gorbachev}},\ }\href {\doibase 10.1103/PhysRevLett.103.226801} {\bibfield
  {journal} {\bibinfo  {journal} {Phys. Rev. Lett.}\ }\textbf {\bibinfo
  {volume} {103}},\ \bibinfo {pages} {226801} (\bibinfo {year}
  {2009})}\BibitemShut {NoStop}%
\bibitem [{\citenamefont {Skakalova}\ and\ \citenamefont
  {Kaiser}(2014)}]{skakalova2014graphene}%
  \BibitemOpen
  \bibfield  {author} {\bibinfo {author} {\bibfnamefont {V.}~\bibnamefont
  {Skakalova}}\ and\ \bibinfo {author} {\bibfnamefont {A.}~\bibnamefont
  {Kaiser}},\ }\href {https://books.google.com.br/books?id=rf6iAgAAQBAJ} {\emph
  {\bibinfo {title} {Graphene: Properties, Preparation, Characterisation and
  Devices}}},\ Woodhead Publishing Series in Electronic and Optical Materials\
  (\bibinfo  {publisher} {Elsevier Science},\ \bibinfo {year}
  {2014})\BibitemShut {NoStop}%
\bibitem [{\citenamefont {Torre}\ \emph
  {et~al.}(2015{\natexlab{a}})\citenamefont {Torre}, \citenamefont {Tomadin},
  \citenamefont {Geim},\ and\ \citenamefont {Polini}}]{PhysRevB.92.165433}%
  \BibitemOpen
  \bibfield  {author} {\bibinfo {author} {\bibfnamefont {I.}~\bibnamefont
  {Torre}}, \bibinfo {author} {\bibfnamefont {A.}~\bibnamefont {Tomadin}},
  \bibinfo {author} {\bibfnamefont {A.~K.}\ \bibnamefont {Geim}}, \ and\
  \bibinfo {author} {\bibfnamefont {M.}~\bibnamefont {Polini}},\ }\href
  {\doibase 10.1103/PhysRevB.92.165433} {\bibfield  {journal} {\bibinfo
  {journal} {Phys. Rev. B}\ }\textbf {\bibinfo {volume} {92}},\ \bibinfo
  {pages} {165433} (\bibinfo {year} {2015}{\natexlab{a}})}\BibitemShut
  {NoStop}%
\bibitem [{\citenamefont {Pellegrino}\ \emph {et~al.}(2016)\citenamefont
  {Pellegrino}, \citenamefont {Torre}, \citenamefont {Geim},\ and\
  \citenamefont {Polini}}]{pellegrino2016electron}%
  \BibitemOpen
  \bibfield  {author} {\bibinfo {author} {\bibfnamefont {F.~M.~D.}\
  \bibnamefont {Pellegrino}}, \bibinfo {author} {\bibfnamefont
  {I.}~\bibnamefont {Torre}}, \bibinfo {author} {\bibfnamefont {A.~K.}\
  \bibnamefont {Geim}}, \ and\ \bibinfo {author} {\bibfnamefont
  {M.}~\bibnamefont {Polini}},\ }\href {\doibase 10.1103/PhysRevB.94.155414}
  {\bibfield  {journal} {\bibinfo  {journal} {Phys. Rev. B}\ }\textbf {\bibinfo
  {volume} {94}},\ \bibinfo {pages} {155414} (\bibinfo {year}
  {2016})}\BibitemShut {NoStop}%
\bibitem [{\citenamefont {Levitov}\ and\ \citenamefont
  {Falkovich}(2016)}]{levitov16}%
  \BibitemOpen
  \bibfield  {author} {\bibinfo {author} {\bibfnamefont {L.}~\bibnamefont
  {Levitov}}\ and\ \bibinfo {author} {\bibfnamefont {G.}~\bibnamefont
  {Falkovich}},\ }\href {http://dx.doi.org/10.1038/nphys3667} {\bibfield
  {journal} {\bibinfo  {journal} {Nature Phys.}\ }\textbf {\bibinfo {volume}
  {12}},\ \bibinfo {pages} {672–} (\bibinfo {year} {2016})}\BibitemShut
  {NoStop}%
\bibitem [{\citenamefont {Guo}\ \emph {et~al.}(2017)\citenamefont {Guo},
  \citenamefont {Ilseven}, \citenamefont {Falkovich},\ and\ \citenamefont
  {Levitov}}]{guo2017higher}%
  \BibitemOpen
  \bibfield  {author} {\bibinfo {author} {\bibfnamefont {H.}~\bibnamefont
  {Guo}}, \bibinfo {author} {\bibfnamefont {E.}~\bibnamefont {Ilseven}},
  \bibinfo {author} {\bibfnamefont {G.}~\bibnamefont {Falkovich}}, \ and\
  \bibinfo {author} {\bibfnamefont {L.~S.}\ \bibnamefont {Levitov}},\ }\href
  {\doibase 10.1073/pnas.1612181114} {\bibfield  {journal} {\bibinfo  {journal}
  {Proceedings of the National Academy of Sciences}\ }\textbf {\bibinfo
  {volume} {114}},\ \bibinfo {pages} {3068} (\bibinfo {year}
  {2017})}\BibitemShut {NoStop}%
\bibitem [{\citenamefont {Krishna~Kumar}\ \emph {et~al.}(2017)\citenamefont
  {Krishna~Kumar}, \citenamefont {Bandurin}, \citenamefont {Pellegrino},
  \citenamefont {Cao}, \citenamefont {Principi}, \citenamefont {Guo},
  \citenamefont {Auton}, \citenamefont {Ben~Shalom}, \citenamefont
  {Ponomarenko}, \citenamefont {Falkovich}, \citenamefont {Watanabe},
  \citenamefont {Taniguchi}, \citenamefont {Grigorieva}, \citenamefont
  {Levitov}, \citenamefont {Polini},\ and\ \citenamefont
  {Geim}}]{kumar2017super}%
  \BibitemOpen
  \bibfield  {author} {\bibinfo {author} {\bibfnamefont {R.}~\bibnamefont
  {Krishna~Kumar}}, \bibinfo {author} {\bibfnamefont {D.~A.}\ \bibnamefont
  {Bandurin}}, \bibinfo {author} {\bibfnamefont {F.~M.~D.}\ \bibnamefont
  {Pellegrino}}, \bibinfo {author} {\bibfnamefont {Y.}~\bibnamefont {Cao}},
  \bibinfo {author} {\bibfnamefont {A.}~\bibnamefont {Principi}}, \bibinfo
  {author} {\bibfnamefont {H.}~\bibnamefont {Guo}}, \bibinfo {author}
  {\bibfnamefont {G.~H.}\ \bibnamefont {Auton}}, \bibinfo {author}
  {\bibfnamefont {M.}~\bibnamefont {Ben~Shalom}}, \bibinfo {author}
  {\bibfnamefont {L.~A.}\ \bibnamefont {Ponomarenko}}, \bibinfo {author}
  {\bibfnamefont {G.}~\bibnamefont {Falkovich}}, \bibinfo {author}
  {\bibfnamefont {K.}~\bibnamefont {Watanabe}}, \bibinfo {author}
  {\bibfnamefont {T.}~\bibnamefont {Taniguchi}}, \bibinfo {author}
  {\bibfnamefont {I.~V.}\ \bibnamefont {Grigorieva}}, \bibinfo {author}
  {\bibfnamefont {L.~S.}\ \bibnamefont {Levitov}}, \bibinfo {author}
  {\bibfnamefont {M.}~\bibnamefont {Polini}}, \ and\ \bibinfo {author}
  {\bibfnamefont {A.~K.}\ \bibnamefont {Geim}},\ }\href@noop {} {\bibfield
  {journal} {\bibinfo  {journal} {Nature Phys.}\ } (\bibinfo {year} {2017})},\
  \Eprint {http://arxiv.org/abs/10.1038/nphys4240} {10.1038/nphys4240}
  \BibitemShut {NoStop}%
\bibitem [{\citenamefont {Crossno}\ \emph {et~al.}(2016)\citenamefont
  {Crossno}, \citenamefont {Shi}, \citenamefont {Wang}, \citenamefont {Liu},
  \citenamefont {Harzheim}, \citenamefont {Lucas}, \citenamefont {Sachdev},
  \citenamefont {Kim}, \citenamefont {Taniguchi}, \citenamefont {Watanabe},
  \citenamefont {Ohki},\ and\ \citenamefont {Fong}}]{crossno16}%
  \BibitemOpen
  \bibfield  {author} {\bibinfo {author} {\bibfnamefont {J.}~\bibnamefont
  {Crossno}}, \bibinfo {author} {\bibfnamefont {J.~K.}\ \bibnamefont {Shi}},
  \bibinfo {author} {\bibfnamefont {K.}~\bibnamefont {Wang}}, \bibinfo {author}
  {\bibfnamefont {X.}~\bibnamefont {Liu}}, \bibinfo {author} {\bibfnamefont
  {A.}~\bibnamefont {Harzheim}}, \bibinfo {author} {\bibfnamefont
  {A.}~\bibnamefont {Lucas}}, \bibinfo {author} {\bibfnamefont
  {S.}~\bibnamefont {Sachdev}}, \bibinfo {author} {\bibfnamefont
  {P.}~\bibnamefont {Kim}}, \bibinfo {author} {\bibfnamefont {T.}~\bibnamefont
  {Taniguchi}}, \bibinfo {author} {\bibfnamefont {K.}~\bibnamefont {Watanabe}},
  \bibinfo {author} {\bibfnamefont {T.~A.}\ \bibnamefont {Ohki}}, \ and\
  \bibinfo {author} {\bibfnamefont {K.~C.}\ \bibnamefont {Fong}},\ }\href
  {\doibase 10.1126/science.aad0343} {\bibfield  {journal} {\bibinfo  {journal}
  {Science}\ }\textbf {\bibinfo {volume} {351}},\ \bibinfo {pages} {1058}
  (\bibinfo {year} {2016})}\BibitemShut {NoStop}%
\bibitem [{\citenamefont {Smyth}\ and\ \citenamefont
  {Moum}(2012)}]{smyth2012ocean}%
  \BibitemOpen
  \bibfield  {author} {\bibinfo {author} {\bibfnamefont {W.~D.}\ \bibnamefont
  {Smyth}}\ and\ \bibinfo {author} {\bibfnamefont {J.~N.}\ \bibnamefont
  {Moum}},\ }\href {http://dx.doi.org/10.5670/oceanog.2012.49} {\bibfield
  {journal} {\bibinfo  {journal} {Oceanography}\ }\textbf {\bibinfo {volume}
  {25}},\ \bibinfo {pages} {140} (\bibinfo {year} {2012})}\BibitemShut
  {NoStop}%
\bibitem [{\citenamefont {Theilhaber}\ and\ \citenamefont
  {Birdsall}(1989)}]{PhysRevLett.62.772}%
  \BibitemOpen
  \bibfield  {author} {\bibinfo {author} {\bibfnamefont {K.}~\bibnamefont
  {Theilhaber}}\ and\ \bibinfo {author} {\bibfnamefont {C.~K.}\ \bibnamefont
  {Birdsall}},\ }\href {\doibase 10.1103/PhysRevLett.62.772} {\bibfield
  {journal} {\bibinfo  {journal} {Phys. Rev. Lett.}\ }\textbf {\bibinfo
  {volume} {62}},\ \bibinfo {pages} {772} (\bibinfo {year} {1989})}\BibitemShut
  {NoStop}%
\bibitem [{\citenamefont {Wyper}\ and\ \citenamefont
  {Pontin}(2013)}]{wyper2013kelvin}%
  \BibitemOpen
  \bibfield  {author} {\bibinfo {author} {\bibfnamefont {P.~F.}\ \bibnamefont
  {Wyper}}\ and\ \bibinfo {author} {\bibfnamefont {D.~I.}\ \bibnamefont
  {Pontin}},\ }\href {\doibase 10.1063/1.4798516} {\bibfield  {journal}
  {\bibinfo  {journal} {Physics of Plasmas}\ }\textbf {\bibinfo {volume}
  {20}},\ \bibinfo {pages} {032117} (\bibinfo {year} {2013})}\BibitemShut
  {NoStop}%
\bibitem [{\citenamefont {{Chandrasekhar}}(1961)}]{chandrasekhar61}%
  \BibitemOpen
  \bibfield  {author} {\bibinfo {author} {\bibfnamefont {S.}~\bibnamefont
  {{Chandrasekhar}}},\ }\href@noop {} {\emph {\bibinfo {title} {International
  Series of Monographs on Physics, Oxford: Clarendon, 1961}}}\ (\bibinfo {year}
  {1961})\BibitemShut {NoStop}%
\bibitem [{\citenamefont {Mohseni}\ \emph {et~al.}(2015)\citenamefont
  {Mohseni}, \citenamefont {Mendoza}, \citenamefont {Succi},\ and\
  \citenamefont {Herrmann}}]{mohseni15}%
  \BibitemOpen
  \bibfield  {author} {\bibinfo {author} {\bibfnamefont {F.}~\bibnamefont
  {Mohseni}}, \bibinfo {author} {\bibfnamefont {M.}~\bibnamefont {Mendoza}},
  \bibinfo {author} {\bibfnamefont {S.}~\bibnamefont {Succi}}, \ and\ \bibinfo
  {author} {\bibfnamefont {H.~J.}\ \bibnamefont {Herrmann}},\ }\href {\doibase
  10.1103/PhysRevE.92.023309} {\bibfield  {journal} {\bibinfo  {journal} {Phys.
  Rev. E}\ }\textbf {\bibinfo {volume} {92}},\ \bibinfo {pages} {023309}
  (\bibinfo {year} {2015})}\BibitemShut {NoStop}%
\bibitem [{\citenamefont {Hasegawa}\ \emph {et~al.}(2004)\citenamefont
  {Hasegawa}, \citenamefont {Fujimoto}, \citenamefont {Phan}, \citenamefont
  {Reme}, \citenamefont {Balogh}, \citenamefont {Dunlop}, \citenamefont
  {Hashimoto},\ and\ \citenamefont {TanDokoro}}]{hasegawa04}%
  \BibitemOpen
  \bibfield  {author} {\bibinfo {author} {\bibfnamefont {H.}~\bibnamefont
  {Hasegawa}}, \bibinfo {author} {\bibfnamefont {M.}~\bibnamefont {Fujimoto}},
  \bibinfo {author} {\bibfnamefont {T.-D.}\ \bibnamefont {Phan}}, \bibinfo
  {author} {\bibfnamefont {H.}~\bibnamefont {Reme}}, \bibinfo {author}
  {\bibfnamefont {A.}~\bibnamefont {Balogh}}, \bibinfo {author} {\bibfnamefont
  {M.}~\bibnamefont {Dunlop}}, \bibinfo {author} {\bibfnamefont
  {C.}~\bibnamefont {Hashimoto}}, \ and\ \bibinfo {author} {\bibfnamefont
  {R.}~\bibnamefont {TanDokoro}},\ }\href {\doibase 10.1038/nature02799}
  {\bibfield  {journal} {\bibinfo  {journal} {Nature}\ }\textbf {\bibinfo
  {volume} {430}},\ \bibinfo {pages} {755} (\bibinfo {year}
  {2004})}\BibitemShut {NoStop}%
\bibitem [{\citenamefont {Blaauwgeers}\ \emph {et~al.}(2002)\citenamefont
  {Blaauwgeers}, \citenamefont {Eltsov}, \citenamefont {Eska}, \citenamefont
  {Finne}, \citenamefont {Haley}, \citenamefont {Krusius}, \citenamefont
  {Ruohio}, \citenamefont {Skrbek},\ and\ \citenamefont
  {Volovik}}]{Blaauwgeers02}%
  \BibitemOpen
  \bibfield  {author} {\bibinfo {author} {\bibfnamefont {R.}~\bibnamefont
  {Blaauwgeers}}, \bibinfo {author} {\bibfnamefont {V.~B.}\ \bibnamefont
  {Eltsov}}, \bibinfo {author} {\bibfnamefont {G.}~\bibnamefont {Eska}},
  \bibinfo {author} {\bibfnamefont {A.~P.}\ \bibnamefont {Finne}}, \bibinfo
  {author} {\bibfnamefont {R.~P.}\ \bibnamefont {Haley}}, \bibinfo {author}
  {\bibfnamefont {M.}~\bibnamefont {Krusius}}, \bibinfo {author} {\bibfnamefont
  {J.~J.}\ \bibnamefont {Ruohio}}, \bibinfo {author} {\bibfnamefont
  {L.}~\bibnamefont {Skrbek}}, \ and\ \bibinfo {author} {\bibfnamefont {G.~E.}\
  \bibnamefont {Volovik}},\ }\href {\doibase 10.1103/PhysRevLett.89.155301}
  {\bibfield  {journal} {\bibinfo  {journal} {Phys. Rev. Lett.}\ }\textbf
  {\bibinfo {volume} {89}},\ \bibinfo {pages} {155301} (\bibinfo {year}
  {2002})}\BibitemShut {NoStop}%
\bibitem [{\citenamefont {Bodo}\ \emph {et~al.}(2004)\citenamefont {Bodo},
  \citenamefont {Mignone},\ and\ \citenamefont {Rosner}}]{bodo04}%
  \BibitemOpen
  \bibfield  {author} {\bibinfo {author} {\bibfnamefont {G.}~\bibnamefont
  {Bodo}}, \bibinfo {author} {\bibfnamefont {A.}~\bibnamefont {Mignone}}, \
  and\ \bibinfo {author} {\bibfnamefont {R.}~\bibnamefont {Rosner}},\ }\href
  {\doibase 10.1103/PhysRevE.70.036304} {\bibfield  {journal} {\bibinfo
  {journal} {Phys. Rev. E}\ }\textbf {\bibinfo {volume} {70}},\ \bibinfo
  {pages} {036304} (\bibinfo {year} {2004})}\BibitemShut {NoStop}%
\bibitem [{\citenamefont {Perucho}\ \emph {et~al.}(2004)\citenamefont
  {Perucho}, \citenamefont {Hanasz}, \citenamefont {Mart{\'\i}},\ and\
  \citenamefont {Sol}}]{perucho04}%
  \BibitemOpen
  \bibfield  {author} {\bibinfo {author} {\bibfnamefont {M.}~\bibnamefont
  {Perucho}}, \bibinfo {author} {\bibfnamefont {M.}~\bibnamefont {Hanasz}},
  \bibinfo {author} {\bibfnamefont {J.-M.}\ \bibnamefont {Mart{\'\i}}}, \ and\
  \bibinfo {author} {\bibfnamefont {H.}~\bibnamefont {Sol}},\ }\href {\doibase
  10.1051/0004--6361:20040349} {\bibfield  {journal} {\bibinfo  {journal}
  {Astronomy \& Astrophysics}\ }\textbf {\bibinfo {volume} {427}},\ \bibinfo
  {pages} {415} (\bibinfo {year} {2004})}\BibitemShut {NoStop}%
\bibitem [{\citenamefont {Coelho}\ \emph {et~al.}(2015)\citenamefont {Coelho},
  \citenamefont {Calvao}, \citenamefont {Reis},\ and\ \citenamefont
  {Siffert}}]{0143-0807-36-1-015007}%
  \BibitemOpen
  \bibfield  {author} {\bibinfo {author} {\bibfnamefont {R.~C.~V.}\
  \bibnamefont {Coelho}}, \bibinfo {author} {\bibfnamefont {M.~O.}\
  \bibnamefont {Calvao}}, \bibinfo {author} {\bibfnamefont {R.~R.~R.}\
  \bibnamefont {Reis}}, \ and\ \bibinfo {author} {\bibfnamefont {B.~B.}\
  \bibnamefont {Siffert}},\ }\href
  {http://stacks.iop.org/0143-0807/36/i=1/a=015007} {\bibfield  {journal}
  {\bibinfo  {journal} {European Journal of Physics}\ }\textbf {\bibinfo
  {volume} {36}},\ \bibinfo {pages} {015007} (\bibinfo {year}
  {2015})}\BibitemShut {NoStop}%
\bibitem [{\citenamefont {Cercignani}\ and\ \citenamefont
  {Kremer}(2002)}]{cercignani02}%
  \BibitemOpen
  \bibfield  {author} {\bibinfo {author} {\bibfnamefont {C.}~\bibnamefont
  {Cercignani}}\ and\ \bibinfo {author} {\bibfnamefont {G.~M.}\ \bibnamefont
  {Kremer}},\ }\enquote {\bibinfo {title} {Relativistic boltzmann equation},}\
  in\ \href@noop {} {\emph {\bibinfo {booktitle} {The Relativistic Boltzmann
  Equation: Theory and Applications}}}\ (\bibinfo  {publisher} {Birkh{\"a}user
  Basel},\ \bibinfo {address} {Basel},\ \bibinfo {year} {2002})\BibitemShut
  {NoStop}%
\bibitem [{\citenamefont {Kremer}(2010)}]{kremer10}%
  \BibitemOpen
  \bibfield  {author} {\bibinfo {author} {\bibfnamefont {G.~M.}\ \bibnamefont
  {Kremer}},\ }\href {\doibase 10.1007/978-3-642-11696-4} {\emph {\bibinfo
  {title} {An Introduction to the Boltzmann Equation and Transport Processes in
  Gases}}},\ edited by\ \bibinfo {editor} {\bibfnamefont {S.-V.~B.}\
  \bibnamefont {Heidelberg}}\ (\bibinfo  {publisher} {Springer-Verlag Berlin
  Heidelberg},\ \bibinfo {year} {2010})\BibitemShut {NoStop}%
\bibitem [{\citenamefont {Narozhny}\ \emph {et~al.}(2017)\citenamefont
  {Narozhny}, \citenamefont {Gornyi}, \citenamefont {Mirlin},\ and\
  \citenamefont {Schmalian}}]{narozhny2017hydrodynamic}%
  \BibitemOpen
  \bibfield  {author} {\bibinfo {author} {\bibfnamefont {B.~N.}\ \bibnamefont
  {Narozhny}}, \bibinfo {author} {\bibfnamefont {I.~V.}\ \bibnamefont
  {Gornyi}}, \bibinfo {author} {\bibfnamefont {A.~D.}\ \bibnamefont {Mirlin}},
  \ and\ \bibinfo {author} {\bibfnamefont {J.}~\bibnamefont {Schmalian}},\
  }\href@noop {} {\bibfield  {journal} {\bibinfo  {journal} {Ann. Phys.
  (Berlin)}\ }\textbf {\bibinfo {volume} {529}},\ \bibinfo {pages} {1700043}
  (\bibinfo {year} {2017})}\BibitemShut {NoStop}%
\bibitem [{\citenamefont {Briskot}\ \emph {et~al.}(2015)\citenamefont
  {Briskot}, \citenamefont {Sch\"utt}, \citenamefont {Gornyi}, \citenamefont
  {Titov}, \citenamefont {Narozhny},\ and\ \citenamefont
  {Mirlin}}]{PhysRevB.92.115426}%
  \BibitemOpen
  \bibfield  {author} {\bibinfo {author} {\bibfnamefont {U.}~\bibnamefont
  {Briskot}}, \bibinfo {author} {\bibfnamefont {M.}~\bibnamefont {Sch\"utt}},
  \bibinfo {author} {\bibfnamefont {I.~V.}\ \bibnamefont {Gornyi}}, \bibinfo
  {author} {\bibfnamefont {M.}~\bibnamefont {Titov}}, \bibinfo {author}
  {\bibfnamefont {B.~N.}\ \bibnamefont {Narozhny}}, \ and\ \bibinfo {author}
  {\bibfnamefont {A.~D.}\ \bibnamefont {Mirlin}},\ }\href {\doibase
  10.1103/PhysRevB.92.115426} {\bibfield  {journal} {\bibinfo  {journal} {Phys.
  Rev. B}\ }\textbf {\bibinfo {volume} {92}},\ \bibinfo {pages} {115426}
  (\bibinfo {year} {2015})}\BibitemShut {NoStop}%
\bibitem [{\citenamefont {Fritz}\ \emph {et~al.}(2008)\citenamefont {Fritz},
  \citenamefont {Schmalian}, \citenamefont {M\"uller},\ and\ \citenamefont
  {Sachdev}}]{PhysRevB.78.085416}%
  \BibitemOpen
  \bibfield  {author} {\bibinfo {author} {\bibfnamefont {L.}~\bibnamefont
  {Fritz}}, \bibinfo {author} {\bibfnamefont {J.}~\bibnamefont {Schmalian}},
  \bibinfo {author} {\bibfnamefont {M.}~\bibnamefont {M\"uller}}, \ and\
  \bibinfo {author} {\bibfnamefont {S.}~\bibnamefont {Sachdev}},\ }\href
  {\doibase 10.1103/PhysRevB.78.085416} {\bibfield  {journal} {\bibinfo
  {journal} {Phys. Rev. B}\ }\textbf {\bibinfo {volume} {78}},\ \bibinfo
  {pages} {085416} (\bibinfo {year} {2008})}\BibitemShut {NoStop}%
\bibitem [{\citenamefont {Govorov}\ and\ \citenamefont
  {Heremans}(2004)}]{PhysRevLett.92.026803}%
  \BibitemOpen
  \bibfield  {author} {\bibinfo {author} {\bibfnamefont {A.~O.}\ \bibnamefont
  {Govorov}}\ and\ \bibinfo {author} {\bibfnamefont {J.~J.}\ \bibnamefont
  {Heremans}},\ }\href {\doibase 10.1103/PhysRevLett.92.026803} {\bibfield
  {journal} {\bibinfo  {journal} {Phys. Rev. Lett.}\ }\textbf {\bibinfo
  {volume} {92}},\ \bibinfo {pages} {026803} (\bibinfo {year}
  {2004})}\BibitemShut {NoStop}%
\bibitem [{\citenamefont {M\"uller}\ \emph {et~al.}(2008)\citenamefont
  {M\"uller}, \citenamefont {Fritz},\ and\ \citenamefont
  {Sachdev}}]{PhysRevB.78.115406}%
  \BibitemOpen
  \bibfield  {author} {\bibinfo {author} {\bibfnamefont {M.}~\bibnamefont
  {M\"uller}}, \bibinfo {author} {\bibfnamefont {L.}~\bibnamefont {Fritz}}, \
  and\ \bibinfo {author} {\bibfnamefont {S.}~\bibnamefont {Sachdev}},\ }\href
  {\doibase 10.1103/PhysRevB.78.115406} {\bibfield  {journal} {\bibinfo
  {journal} {Phys. Rev. B}\ }\textbf {\bibinfo {volume} {78}},\ \bibinfo
  {pages} {115406} (\bibinfo {year} {2008})}\BibitemShut {NoStop}%
\bibitem [{\citenamefont {Narozhny}\ \emph {et~al.}(2012)\citenamefont
  {Narozhny}, \citenamefont {Titov}, \citenamefont {Gornyi},\ and\
  \citenamefont {Ostrovsky}}]{PhysRevB.85.195421}%
  \BibitemOpen
  \bibfield  {author} {\bibinfo {author} {\bibfnamefont {B.~N.}\ \bibnamefont
  {Narozhny}}, \bibinfo {author} {\bibfnamefont {M.}~\bibnamefont {Titov}},
  \bibinfo {author} {\bibfnamefont {I.~V.}\ \bibnamefont {Gornyi}}, \ and\
  \bibinfo {author} {\bibfnamefont {P.~M.}\ \bibnamefont {Ostrovsky}},\ }\href
  {\doibase 10.1103/PhysRevB.85.195421} {\bibfield  {journal} {\bibinfo
  {journal} {Phys. Rev. B}\ }\textbf {\bibinfo {volume} {85}},\ \bibinfo
  {pages} {195421} (\bibinfo {year} {2012})}\BibitemShut {NoStop}%
\bibitem [{\citenamefont {Narozhny}\ \emph {et~al.}(2015)\citenamefont
  {Narozhny}, \citenamefont {Gornyi}, \citenamefont {Titov}, \citenamefont
  {Sch\"utt},\ and\ \citenamefont {Mirlin}}]{PhysRevB.91.035414}%
  \BibitemOpen
  \bibfield  {author} {\bibinfo {author} {\bibfnamefont {B.~N.}\ \bibnamefont
  {Narozhny}}, \bibinfo {author} {\bibfnamefont {I.~V.}\ \bibnamefont
  {Gornyi}}, \bibinfo {author} {\bibfnamefont {M.}~\bibnamefont {Titov}},
  \bibinfo {author} {\bibfnamefont {M.}~\bibnamefont {Sch\"utt}}, \ and\
  \bibinfo {author} {\bibfnamefont {A.~D.}\ \bibnamefont {Mirlin}},\ }\href
  {\doibase 10.1103/PhysRevB.91.035414} {\bibfield  {journal} {\bibinfo
  {journal} {Phys. Rev. B}\ }\textbf {\bibinfo {volume} {91}},\ \bibinfo
  {pages} {035414} (\bibinfo {year} {2015})}\BibitemShut {NoStop}%
\bibitem [{\citenamefont {Principi}\ \emph {et~al.}(2016)\citenamefont
  {Principi}, \citenamefont {Vignale}, \citenamefont {Carrega},\ and\
  \citenamefont {Polini}}]{PhysRevB.93.125410}%
  \BibitemOpen
  \bibfield  {author} {\bibinfo {author} {\bibfnamefont {A.}~\bibnamefont
  {Principi}}, \bibinfo {author} {\bibfnamefont {G.}~\bibnamefont {Vignale}},
  \bibinfo {author} {\bibfnamefont {M.}~\bibnamefont {Carrega}}, \ and\
  \bibinfo {author} {\bibfnamefont {M.}~\bibnamefont {Polini}},\ }\href
  {\doibase 10.1103/PhysRevB.93.125410} {\bibfield  {journal} {\bibinfo
  {journal} {Phys. Rev. B}\ }\textbf {\bibinfo {volume} {93}},\ \bibinfo
  {pages} {125410} (\bibinfo {year} {2016})}\BibitemShut {NoStop}%
\bibitem [{\citenamefont {Chapman}\ and\ \citenamefont
  {Cowling}(1970)}]{chapman70}%
  \BibitemOpen
  \bibfield  {author} {\bibinfo {author} {\bibfnamefont {S.}~\bibnamefont
  {Chapman}}\ and\ \bibinfo {author} {\bibfnamefont {T.~G.}\ \bibnamefont
  {Cowling}},\ }\href@noop {} {\emph {\bibinfo {title} {The mathematical theory
  of non-uniform gases: an account of the kinetic theory of viscosity, thermal
  conduction and diffusion in gases}}}\ (\bibinfo  {publisher} {Cambridge
  university press},\ \bibinfo {year} {1970})\BibitemShut {NoStop}%
\bibitem [{\citenamefont {Kr{\"u}ger}\ \emph {et~al.}(2016)\citenamefont
  {Kr{\"u}ger}, \citenamefont {Kusumaatmaja}, \citenamefont {Kuzmin},
  \citenamefont {Shardt}, \citenamefont {Silva},\ and\ \citenamefont
  {Viggen}}]{kruger16}%
  \BibitemOpen
  \bibfield  {author} {\bibinfo {author} {\bibfnamefont {T.}~\bibnamefont
  {Kr{\"u}ger}}, \bibinfo {author} {\bibfnamefont {H.}~\bibnamefont
  {Kusumaatmaja}}, \bibinfo {author} {\bibfnamefont {A.}~\bibnamefont
  {Kuzmin}}, \bibinfo {author} {\bibfnamefont {O.}~\bibnamefont {Shardt}},
  \bibinfo {author} {\bibfnamefont {G.}~\bibnamefont {Silva}}, \ and\ \bibinfo
  {author} {\bibfnamefont {E.}~\bibnamefont {Viggen}},\ }\href {\doibase
  10.1007/978-3-319-44649-3} {\emph {\bibinfo {title} {The Lattice Boltzmann
  Method: Principles and Practice}}},\ Graduate Texts in Physics\ (\bibinfo
  {publisher} {Springer International Publishing},\ \bibinfo {year}
  {2016})\BibitemShut {NoStop}%
\bibitem [{\citenamefont {Succi}(2001)}]{succi01}%
  \BibitemOpen
  \bibfield  {author} {\bibinfo {author} {\bibfnamefont {S.}~\bibnamefont
  {Succi}},\ }\href {http://ukcatalogue.oup.com/product/9780198503989.do}
  {\emph {\bibinfo {title} {The Lattice Boltzmann Equation for Fluid Dynamics
  and Beyond}}},\ edited by\ \bibinfo {editor} {\bibfnamefont {O.~U.}\
  \bibnamefont {Press}}\ (\bibinfo  {publisher} {Clarendon Press},\ \bibinfo
  {year} {2001})\BibitemShut {NoStop}%
\bibitem [{\citenamefont {Coelho}\ \emph {et~al.}(2014)\citenamefont {Coelho},
  \citenamefont {Ilha}, \citenamefont {Doria}, \citenamefont {Pereira},\ and\
  \citenamefont {Aibe}}]{coelho14}%
  \BibitemOpen
  \bibfield  {author} {\bibinfo {author} {\bibfnamefont {R.~C.~V.}\
  \bibnamefont {Coelho}}, \bibinfo {author} {\bibfnamefont {A.}~\bibnamefont
  {Ilha}}, \bibinfo {author} {\bibfnamefont {M.~M.}\ \bibnamefont {Doria}},
  \bibinfo {author} {\bibfnamefont {R.~M.}\ \bibnamefont {Pereira}}, \ and\
  \bibinfo {author} {\bibfnamefont {V.~Y.}\ \bibnamefont {Aibe}},\ }\href
  {\doibase 10.1103/PhysRevE.89.043302} {\bibfield  {journal} {\bibinfo
  {journal} {Phys. Rev. E}\ }\textbf {\bibinfo {volume} {89}},\ \bibinfo
  {pages} {043302} (\bibinfo {year} {2014})}\BibitemShut {NoStop}%
\bibitem [{\citenamefont {Coelho}\ \emph {et~al.}(2016)\citenamefont {Coelho},
  \citenamefont {Ilha},\ and\ \citenamefont {Doria}}]{coelho16-2}%
  \BibitemOpen
  \bibfield  {author} {\bibinfo {author} {\bibfnamefont {R.~C.~V.}\
  \bibnamefont {Coelho}}, \bibinfo {author} {\bibfnamefont {A.~S.}\
  \bibnamefont {Ilha}}, \ and\ \bibinfo {author} {\bibfnamefont {M.~M.}\
  \bibnamefont {Doria}},\ }\href
  {http://stacks.iop.org/0295-5075/116/i=2/a=20001} {\bibfield  {journal}
  {\bibinfo  {journal} {EPL (Europhysics Letters)}\ }\textbf {\bibinfo {volume}
  {116}},\ \bibinfo {pages} {20001} (\bibinfo {year} {2016})}\BibitemShut
  {NoStop}%
\bibitem [{\citenamefont {Yang}\ and\ \citenamefont {Hung}(2009)}]{yang09}%
  \BibitemOpen
  \bibfield  {author} {\bibinfo {author} {\bibfnamefont {J.-Y.}\ \bibnamefont
  {Yang}}\ and\ \bibinfo {author} {\bibfnamefont {L.-H.}\ \bibnamefont
  {Hung}},\ }\href {\doibase 10.1103/PhysRevE.79.056708} {\bibfield  {journal}
  {\bibinfo  {journal} {Phys. Rev. E}\ }\textbf {\bibinfo {volume} {79}},\
  \bibinfo {pages} {056708} (\bibinfo {year} {2009})}\BibitemShut {NoStop}%
\bibitem [{\citenamefont {Palpacelli}\ and\ \citenamefont
  {Succi}(2008)}]{palpacelli2008quantum}%
  \BibitemOpen
  \bibfield  {author} {\bibinfo {author} {\bibfnamefont {S.}~\bibnamefont
  {Palpacelli}}\ and\ \bibinfo {author} {\bibfnamefont {S.}~\bibnamefont
  {Succi}},\ }\href@noop {} {\bibfield  {journal} {\bibinfo  {journal}
  {Communications in Computational Physics}\ }\textbf {\bibinfo {volume} {4}},\
  \bibinfo {pages} {980} (\bibinfo {year} {2008})}\BibitemShut {NoStop}%
\bibitem [{\citenamefont {Palpacelli}\ \emph {et~al.}(2007)\citenamefont
  {Palpacelli}, \citenamefont {Succi},\ and\ \citenamefont
  {Spigler}}]{PhysRevE.76.036712}%
  \BibitemOpen
  \bibfield  {author} {\bibinfo {author} {\bibfnamefont {S.}~\bibnamefont
  {Palpacelli}}, \bibinfo {author} {\bibfnamefont {S.}~\bibnamefont {Succi}}, \
  and\ \bibinfo {author} {\bibfnamefont {R.}~\bibnamefont {Spigler}},\ }\href
  {\doibase 10.1103/PhysRevE.76.036712} {\bibfield  {journal} {\bibinfo
  {journal} {Phys. Rev. E}\ }\textbf {\bibinfo {volume} {76}},\ \bibinfo
  {pages} {036712} (\bibinfo {year} {2007})}\BibitemShut {NoStop}%
\bibitem [{\citenamefont {{Solorzano}}\ \emph {et~al.}(2017)\citenamefont
  {{Solorzano}}, \citenamefont {{Mendoza}}, \citenamefont {{Succi}},\ and\
  \citenamefont {{Herrmann}}}]{solorzano2017lattice}%
  \BibitemOpen
  \bibfield  {author} {\bibinfo {author} {\bibfnamefont {S.}~\bibnamefont
  {{Solorzano}}}, \bibinfo {author} {\bibfnamefont {M.}~\bibnamefont
  {{Mendoza}}}, \bibinfo {author} {\bibfnamefont {S.}~\bibnamefont {{Succi}}},
  \ and\ \bibinfo {author} {\bibfnamefont {H.}~\bibnamefont {{Herrmann}}},\
  }\href@noop {} {\bibfield  {journal} {\bibinfo  {journal} {ArXiv e-prints}\ }
  (\bibinfo {year} {2017})},\ \Eprint {http://arxiv.org/abs/1709.05934}
  {arXiv:1709.05934 [physics.comp-ph]} \BibitemShut {NoStop}%
\bibitem [{\citenamefont {Mendoza}\ \emph {et~al.}(2010)\citenamefont
  {Mendoza}, \citenamefont {Boghosian}, \citenamefont {Herrmann},\ and\
  \citenamefont {Succi}}]{mendoza10}%
  \BibitemOpen
  \bibfield  {author} {\bibinfo {author} {\bibfnamefont {M.}~\bibnamefont
  {Mendoza}}, \bibinfo {author} {\bibfnamefont {B.~M.}\ \bibnamefont
  {Boghosian}}, \bibinfo {author} {\bibfnamefont {H.~J.}\ \bibnamefont
  {Herrmann}}, \ and\ \bibinfo {author} {\bibfnamefont {S.}~\bibnamefont
  {Succi}},\ }\href {\doibase 10.1103/PhysRevLett.105.014502} {\bibfield
  {journal} {\bibinfo  {journal} {Phys. Rev. Lett.}\ }\textbf {\bibinfo
  {volume} {105}},\ \bibinfo {pages} {014502} (\bibinfo {year}
  {2010})}\BibitemShut {NoStop}%
\bibitem [{\citenamefont {Gabbana}\ \emph {et~al.}(2017)\citenamefont
  {Gabbana}, \citenamefont {Mendoza}, \citenamefont {Succi},\ and\
  \citenamefont {Tripiccione}}]{PhysRevE.95.053304}%
  \BibitemOpen
  \bibfield  {author} {\bibinfo {author} {\bibfnamefont {A.}~\bibnamefont
  {Gabbana}}, \bibinfo {author} {\bibfnamefont {M.}~\bibnamefont {Mendoza}},
  \bibinfo {author} {\bibfnamefont {S.}~\bibnamefont {Succi}}, \ and\ \bibinfo
  {author} {\bibfnamefont {R.}~\bibnamefont {Tripiccione}},\ }\href {\doibase
  10.1103/PhysRevE.95.053304} {\bibfield  {journal} {\bibinfo  {journal} {Phys.
  Rev. E}\ }\textbf {\bibinfo {volume} {95}},\ \bibinfo {pages} {053304}
  (\bibinfo {year} {2017})}\BibitemShut {NoStop}%
\bibitem [{\citenamefont {Mendoza}\ \emph {et~al.}(2011)\citenamefont
  {Mendoza}, \citenamefont {Herrmann},\ and\ \citenamefont
  {Succi}}]{mendoza11}%
  \BibitemOpen
  \bibfield  {author} {\bibinfo {author} {\bibfnamefont {M.}~\bibnamefont
  {Mendoza}}, \bibinfo {author} {\bibfnamefont {H.~J.}\ \bibnamefont
  {Herrmann}}, \ and\ \bibinfo {author} {\bibfnamefont {S.}~\bibnamefont
  {Succi}},\ }\href {\doibase 10.1103/PhysRevLett.106.156601} {\bibfield
  {journal} {\bibinfo  {journal} {Phys. Rev. Lett.}\ }\textbf {\bibinfo
  {volume} {106}},\ \bibinfo {pages} {156601} (\bibinfo {year}
  {2011})}\BibitemShut {NoStop}%
\bibitem [{\citenamefont {Oettinger}\ \emph {et~al.}(2013)\citenamefont
  {Oettinger}, \citenamefont {Mendoza},\ and\ \citenamefont
  {Herrmann}}]{oettinger13}%
  \BibitemOpen
  \bibfield  {author} {\bibinfo {author} {\bibfnamefont {D.}~\bibnamefont
  {Oettinger}}, \bibinfo {author} {\bibfnamefont {M.}~\bibnamefont {Mendoza}},
  \ and\ \bibinfo {author} {\bibfnamefont {H.~J.}\ \bibnamefont {Herrmann}},\
  }\href {\doibase 10.1103/PhysRevE.88.013302} {\bibfield  {journal} {\bibinfo
  {journal} {Phys. Rev. E}\ }\textbf {\bibinfo {volume} {88}},\ \bibinfo
  {pages} {013302} (\bibinfo {year} {2013})}\BibitemShut {NoStop}%
\bibitem [{\citenamefont {Furtmaier}\ \emph {et~al.}(2015)\citenamefont
  {Furtmaier}, \citenamefont {Mendoza}, \citenamefont {Karlin}, \citenamefont
  {Succi},\ and\ \citenamefont {Herrmann}}]{furtmaier15}%
  \BibitemOpen
  \bibfield  {author} {\bibinfo {author} {\bibfnamefont {O.}~\bibnamefont
  {Furtmaier}}, \bibinfo {author} {\bibfnamefont {M.}~\bibnamefont {Mendoza}},
  \bibinfo {author} {\bibfnamefont {I.}~\bibnamefont {Karlin}}, \bibinfo
  {author} {\bibfnamefont {S.}~\bibnamefont {Succi}}, \ and\ \bibinfo {author}
  {\bibfnamefont {H.~J.}\ \bibnamefont {Herrmann}},\ }\href {\doibase
  10.1103/PhysRevB.91.085401} {\bibfield  {journal} {\bibinfo  {journal} {Phys.
  Rev. B}\ }\textbf {\bibinfo {volume} {91}},\ \bibinfo {pages} {085401}
  (\bibinfo {year} {2015})}\BibitemShut {NoStop}%
\bibitem [{\citenamefont {Mendoza}\ \emph
  {et~al.}(2013{\natexlab{a}})\citenamefont {Mendoza}, \citenamefont
  {Herrmann},\ and\ \citenamefont {Succi}}]{mendoza13}%
  \BibitemOpen
  \bibfield  {author} {\bibinfo {author} {\bibfnamefont {M.}~\bibnamefont
  {Mendoza}}, \bibinfo {author} {\bibfnamefont {H.~J.}\ \bibnamefont
  {Herrmann}}, \ and\ \bibinfo {author} {\bibfnamefont {S.}~\bibnamefont
  {Succi}},\ }\href {http://dx.doi.org/10.1038/srep01052} {\bibfield  {journal}
  {\bibinfo  {journal} {Scientific Reports}\ }\textbf {\bibinfo {volume} {3}},\
  \bibinfo {pages} {1052} (\bibinfo {year} {2013}{\natexlab{a}})}\BibitemShut
  {NoStop}%
\bibitem [{\citenamefont {{Giordanelli}}\ \emph {et~al.}(2017)\citenamefont
  {{Giordanelli}}, \citenamefont {{Mendoza}},\ and\ \citenamefont
  {{Herrmann}}}]{giordanelli2017modelling}%
  \BibitemOpen
  \bibfield  {author} {\bibinfo {author} {\bibfnamefont {I.}~\bibnamefont
  {{Giordanelli}}}, \bibinfo {author} {\bibfnamefont {M.}~\bibnamefont
  {{Mendoza}}}, \ and\ \bibinfo {author} {\bibfnamefont {H.}~\bibnamefont
  {{Herrmann}}},\ }\href@noop {} {\bibfield  {journal} {\bibinfo  {journal}
  {ArXiv e-prints}\ } (\bibinfo {year} {2017})},\ \Eprint
  {http://arxiv.org/abs/1702.04156} {arXiv:1702.04156 [cond-mat.mtrl-sci]}
  \BibitemShut {NoStop}%
\bibitem [{\citenamefont {Hupp}\ \emph {et~al.}(2011)\citenamefont {Hupp},
  \citenamefont {Mendoza}, \citenamefont {Bouras}, \citenamefont {Succi},\ and\
  \citenamefont {Herrmann}}]{hupp11}%
  \BibitemOpen
  \bibfield  {author} {\bibinfo {author} {\bibfnamefont {D.}~\bibnamefont
  {Hupp}}, \bibinfo {author} {\bibfnamefont {M.}~\bibnamefont {Mendoza}},
  \bibinfo {author} {\bibfnamefont {I.}~\bibnamefont {Bouras}}, \bibinfo
  {author} {\bibfnamefont {S.}~\bibnamefont {Succi}}, \ and\ \bibinfo {author}
  {\bibfnamefont {H.~J.}\ \bibnamefont {Herrmann}},\ }\href {\doibase
  10.1103/PhysRevD.84.125015} {\bibfield  {journal} {\bibinfo  {journal} {Phys.
  Rev. D}\ }\textbf {\bibinfo {volume} {84}},\ \bibinfo {pages} {125015}
  (\bibinfo {year} {2011})}\BibitemShut {NoStop}%
\bibitem [{\citenamefont {Romatschke}\ \emph {et~al.}(2011)\citenamefont
  {Romatschke}, \citenamefont {Mendoza},\ and\ \citenamefont
  {Succi}}]{romatschke11}%
  \BibitemOpen
  \bibfield  {author} {\bibinfo {author} {\bibfnamefont {P.}~\bibnamefont
  {Romatschke}}, \bibinfo {author} {\bibfnamefont {M.}~\bibnamefont {Mendoza}},
  \ and\ \bibinfo {author} {\bibfnamefont {S.}~\bibnamefont {Succi}},\ }\href
  {\doibase 10.1103/PhysRevC.84.034903} {\bibfield  {journal} {\bibinfo
  {journal} {Phys. Rev. C}\ }\textbf {\bibinfo {volume} {84}},\ \bibinfo
  {pages} {034903} (\bibinfo {year} {2011})}\BibitemShut {NoStop}%
\bibitem [{\citenamefont {Mendoza}\ \emph
  {et~al.}(2013{\natexlab{b}})\citenamefont {Mendoza}, \citenamefont {Karlin},
  \citenamefont {Succi},\ and\ \citenamefont {Herrmann}}]{mendoza13-3}%
  \BibitemOpen
  \bibfield  {author} {\bibinfo {author} {\bibfnamefont {M.}~\bibnamefont
  {Mendoza}}, \bibinfo {author} {\bibfnamefont {I.}~\bibnamefont {Karlin}},
  \bibinfo {author} {\bibfnamefont {S.}~\bibnamefont {Succi}}, \ and\ \bibinfo
  {author} {\bibfnamefont {H.~J.}\ \bibnamefont {Herrmann}},\ }\href {\doibase
  10.1103/PhysRevD.87.065027} {\bibfield  {journal} {\bibinfo  {journal} {Phys.
  Rev. D}\ }\textbf {\bibinfo {volume} {87}},\ \bibinfo {pages} {065027}
  (\bibinfo {year} {2013}{\natexlab{b}})}\BibitemShut {NoStop}%
\bibitem [{\citenamefont {Hwa}\ and\ \citenamefont
  {Wang}(2010)}]{hwa2010quark}%
  \BibitemOpen
  \bibfield  {author} {\bibinfo {author} {\bibfnamefont {R.~C.}\ \bibnamefont
  {Hwa}}\ and\ \bibinfo {author} {\bibfnamefont {X.-N.}\ \bibnamefont {Wang}},\
  }\href@noop {} {\emph {\bibinfo {title} {Quark-gluon plasma 4}}}\ (\bibinfo
  {publisher} {World Scientific},\ \bibinfo {year} {2010})\BibitemShut
  {NoStop}%
\bibitem [{\citenamefont {Teaney}(2009)}]{teaney09}%
  \BibitemOpen
  \bibfield  {author} {\bibinfo {author} {\bibfnamefont {D.~A.}\ \bibnamefont
  {Teaney}},\ }\href@noop {} {\bibfield  {journal} {\bibinfo  {journal}
  {Quark-gluon plasma}\ }\textbf {\bibinfo {volume} {4}},\ \bibinfo {pages}
  {207} (\bibinfo {year} {2009})}\BibitemShut {NoStop}%
\bibitem [{\citenamefont {Landau}\ and\ \citenamefont
  {Lifshitz}(2000)}]{landau86}%
  \BibitemOpen
  \bibfield  {author} {\bibinfo {author} {\bibfnamefont {L.~D.}\ \bibnamefont
  {Landau}}\ and\ \bibinfo {author} {\bibfnamefont {E.~M.}\ \bibnamefont
  {Lifshitz}},\ }\href@noop {} {\emph {\bibinfo {title} {Fluid Mechanics}}}\
  (\bibinfo  {publisher} {Butterworth-Heinemann, Oxford, UK},\ \bibinfo {year}
  {2000})\BibitemShut {NoStop}%
\bibitem [{\citenamefont {Mendoza}\ \emph
  {et~al.}(2013{\natexlab{c}})\citenamefont {Mendoza}, \citenamefont {Karlin},
  \citenamefont {Succi},\ and\ \citenamefont {Herrmann}}]{mendoza13-2}%
  \BibitemOpen
  \bibfield  {author} {\bibinfo {author} {\bibfnamefont {M.}~\bibnamefont
  {Mendoza}}, \bibinfo {author} {\bibfnamefont {I.}~\bibnamefont {Karlin}},
  \bibinfo {author} {\bibfnamefont {S.}~\bibnamefont {Succi}}, \ and\ \bibinfo
  {author} {\bibfnamefont {H.~J.}\ \bibnamefont {Herrmann}},\ }\href
  {http://stacks.iop.org/1742-5468/2013/i=02/a=P02036} {\bibfield  {journal}
  {\bibinfo  {journal} {Journal of Statistical Mechanics: Theory and
  Experiment}\ }\textbf {\bibinfo {volume} {2013}},\ \bibinfo {pages} {P02036}
  (\bibinfo {year} {2013}{\natexlab{c}})}\BibitemShut {NoStop}%
\bibitem [{Note1()}]{Note1}%
  \BibitemOpen
  \bibinfo {note} {See Supplemental Material at [URL will be inserted by
  publisher] for more details about the model, which includes the Refs.~\cite
  {mendoza10, kruger16, muller09, coelho14, doria17, coelho16-2,
  cercignani02}.}\BibitemShut {Stop}%
\bibitem [{\citenamefont {{Doria}}\ and\ \citenamefont
  {{Coelho}}(2017)}]{doria17}%
  \BibitemOpen
  \bibfield  {author} {\bibinfo {author} {\bibfnamefont {M.~M.}\ \bibnamefont
  {{Doria}}}\ and\ \bibinfo {author} {\bibfnamefont {R.~C.~V.}\ \bibnamefont
  {{Coelho}}},\ }\href@noop {} {\bibfield  {journal} {\bibinfo  {journal}
  {ArXiv e-prints}\ } (\bibinfo {year} {2017})},\ \Eprint
  {http://arxiv.org/abs/1703.08670} {arXiv:1703.08670 [math-ph]} \BibitemShut
  {NoStop}%
\bibitem [{\citenamefont {Coelho}\ and\ \citenamefont
  {Neumann}(2016)}]{coelho16}%
  \BibitemOpen
  \bibfield  {author} {\bibinfo {author} {\bibfnamefont {R.~C.~V.}\
  \bibnamefont {Coelho}}\ and\ \bibinfo {author} {\bibfnamefont {R.~F.}\
  \bibnamefont {Neumann}},\ }\href
  {http://stacks.iop.org/0143-0807/37/i=5/a=055102} {\bibfield  {journal}
  {\bibinfo  {journal} {European Journal of Physics}\ }\textbf {\bibinfo
  {volume} {37}},\ \bibinfo {pages} {055102} (\bibinfo {year}
  {2016})}\BibitemShut {NoStop}%
\bibitem [{\citenamefont {Mei}\ \emph {et~al.}(2006)\citenamefont {Mei},
  \citenamefont {Luo}, \citenamefont {Lallemand},\ and\ \citenamefont
  {d’Humi{\`e}res}}]{mei06}%
  \BibitemOpen
  \bibfield  {author} {\bibinfo {author} {\bibfnamefont {R.}~\bibnamefont
  {Mei}}, \bibinfo {author} {\bibfnamefont {L.-S.}\ \bibnamefont {Luo}},
  \bibinfo {author} {\bibfnamefont {P.}~\bibnamefont {Lallemand}}, \ and\
  \bibinfo {author} {\bibfnamefont {D.}~\bibnamefont {d’Humi{\`e}res}},\
  }\href@noop {} {\bibfield  {journal} {\bibinfo  {journal} {Computers \&
  Fluids}\ }\textbf {\bibinfo {volume} {35}},\ \bibinfo {pages} {855} (\bibinfo
  {year} {2006})}\BibitemShut {NoStop}%
\bibitem [{\citenamefont {Lallemand}\ and\ \citenamefont
  {Luo}(2000)}]{PhysRevE.61.6546}%
  \BibitemOpen
  \bibfield  {author} {\bibinfo {author} {\bibfnamefont {P.}~\bibnamefont
  {Lallemand}}\ and\ \bibinfo {author} {\bibfnamefont {L.-S.}\ \bibnamefont
  {Luo}},\ }\href {\doibase 10.1103/PhysRevE.61.6546} {\bibfield  {journal}
  {\bibinfo  {journal} {Phys. Rev. E}\ }\textbf {\bibinfo {volume} {61}},\
  \bibinfo {pages} {6546} (\bibinfo {year} {2000})}\BibitemShut {NoStop}%
\bibitem [{\citenamefont {Wu}\ and\ \citenamefont {Shu}(2011)}]{Wu20112246}%
  \BibitemOpen
  \bibfield  {author} {\bibinfo {author} {\bibfnamefont {J.}~\bibnamefont
  {Wu}}\ and\ \bibinfo {author} {\bibfnamefont {C.}~\bibnamefont {Shu}},\
  }\href {\doibase http://dx.doi.org/10.1016/j.jcp.2010.12.013} {\bibfield
  {journal} {\bibinfo  {journal} {Journal of Computational Physics}\ }\textbf
  {\bibinfo {volume} {230}},\ \bibinfo {pages} {2246 } (\bibinfo {year}
  {2011})}\BibitemShut {NoStop}%
\bibitem [{\citenamefont {Yu}\ \emph {et~al.}(2003)\citenamefont {Yu},
  \citenamefont {Mei}, \citenamefont {Luo},\ and\ \citenamefont
  {Shyy}}]{Yu2003329}%
  \BibitemOpen
  \bibfield  {author} {\bibinfo {author} {\bibfnamefont {D.}~\bibnamefont
  {Yu}}, \bibinfo {author} {\bibfnamefont {R.}~\bibnamefont {Mei}}, \bibinfo
  {author} {\bibfnamefont {L.-S.}\ \bibnamefont {Luo}}, \ and\ \bibinfo
  {author} {\bibfnamefont {W.}~\bibnamefont {Shyy}},\ }\href {\doibase
  http://dx.doi.org/10.1016/S0376-0421(03)00003-4} {\bibfield  {journal}
  {\bibinfo  {journal} {Progress in Aerospace Sciences}\ }\textbf {\bibinfo
  {volume} {39}},\ \bibinfo {pages} {329 } (\bibinfo {year}
  {2003})}\BibitemShut {NoStop}%
\bibitem [{\citenamefont {Gan}\ \emph {et~al.}(2011)\citenamefont {Gan},
  \citenamefont {Xu}, \citenamefont {Zhang},\ and\ \citenamefont {Li}}]{gan11}%
  \BibitemOpen
  \bibfield  {author} {\bibinfo {author} {\bibfnamefont {Y.}~\bibnamefont
  {Gan}}, \bibinfo {author} {\bibfnamefont {A.}~\bibnamefont {Xu}}, \bibinfo
  {author} {\bibfnamefont {G.}~\bibnamefont {Zhang}}, \ and\ \bibinfo {author}
  {\bibfnamefont {Y.}~\bibnamefont {Li}},\ }\href {\doibase
  10.1103/PhysRevE.83.056704} {\bibfield  {journal} {\bibinfo  {journal} {Phys.
  Rev. E}\ }\textbf {\bibinfo {volume} {83}},\ \bibinfo {pages} {056704}
  (\bibinfo {year} {2011})}\BibitemShut {NoStop}%
\bibitem [{\citenamefont {Nakayama}(1990)}]{nakayama90}%
  \BibitemOpen
  \bibfield  {author} {\bibinfo {author} {\bibfnamefont {K.}~\bibnamefont
  {Nakayama}},\ }\href@noop {} {\bibfield  {journal} {\bibinfo  {journal}
  {Publications of the Astronomical Society of Japan}\ }\textbf {\bibinfo
  {volume} {42}},\ \bibinfo {pages} {331} (\bibinfo {year} {1990})}\BibitemShut
  {NoStop}%
\bibitem [{\citenamefont {Barreiro}\ \emph {et~al.}(2009)\citenamefont
  {Barreiro}, \citenamefont {Lazzeri}, \citenamefont {Moser}, \citenamefont
  {Mauri},\ and\ \citenamefont {Bachtold}}]{barreiro09}%
  \BibitemOpen
  \bibfield  {author} {\bibinfo {author} {\bibfnamefont {A.}~\bibnamefont
  {Barreiro}}, \bibinfo {author} {\bibfnamefont {M.}~\bibnamefont {Lazzeri}},
  \bibinfo {author} {\bibfnamefont {J.}~\bibnamefont {Moser}}, \bibinfo
  {author} {\bibfnamefont {F.}~\bibnamefont {Mauri}}, \ and\ \bibinfo {author}
  {\bibfnamefont {A.}~\bibnamefont {Bachtold}},\ }\href {\doibase
  10.1103/PhysRevLett.103.076601} {\bibfield  {journal} {\bibinfo  {journal}
  {Phys. Rev. Lett.}\ }\textbf {\bibinfo {volume} {103}},\ \bibinfo {pages}
  {076601} (\bibinfo {year} {2009})}\BibitemShut {NoStop}%
\bibitem [{\citenamefont {Torre}\ \emph
  {et~al.}(2015{\natexlab{b}})\citenamefont {Torre}, \citenamefont {Tomadin},
  \citenamefont {Krahne}, \citenamefont {Pellegrini},\ and\ \citenamefont
  {Polini}}]{torre15}%
  \BibitemOpen
  \bibfield  {author} {\bibinfo {author} {\bibfnamefont {I.}~\bibnamefont
  {Torre}}, \bibinfo {author} {\bibfnamefont {A.}~\bibnamefont {Tomadin}},
  \bibinfo {author} {\bibfnamefont {R.}~\bibnamefont {Krahne}}, \bibinfo
  {author} {\bibfnamefont {V.}~\bibnamefont {Pellegrini}}, \ and\ \bibinfo
  {author} {\bibfnamefont {M.}~\bibnamefont {Polini}},\ }\href {\doibase
  10.1103/PhysRevB.91.081402} {\bibfield  {journal} {\bibinfo  {journal} {Phys.
  Rev. B}\ }\textbf {\bibinfo {volume} {91}},\ \bibinfo {pages} {081402}
  (\bibinfo {year} {2015}{\natexlab{b}})}\BibitemShut {NoStop}%
\bibitem [{\citenamefont {Wehling}\ \emph {et~al.}(2014)\citenamefont
  {Wehling}, \citenamefont {Black-Schaffer},\ and\ \citenamefont
  {Balatsky}}]{wehling14}%
  \BibitemOpen
  \bibfield  {author} {\bibinfo {author} {\bibfnamefont {T.}~\bibnamefont
  {Wehling}}, \bibinfo {author} {\bibfnamefont {A.}~\bibnamefont
  {Black-Schaffer}}, \ and\ \bibinfo {author} {\bibfnamefont {A.}~\bibnamefont
  {Balatsky}},\ }\href {\doibase 10.1080/00018732.2014.927109} {\bibfield
  {journal} {\bibinfo  {journal} {Advances in Physics}\ }\textbf {\bibinfo
  {volume} {63}},\ \bibinfo {pages} {1} (\bibinfo {year} {2014})}\BibitemShut
  {NoStop}%
\bibitem [{\citenamefont {Chan}\ \emph {et~al.}(2016)\citenamefont {Chan},
  \citenamefont {Kvorning}, \citenamefont {Ryu},\ and\ \citenamefont
  {Fradkin}}]{PhysRevB.93.155122}%
  \BibitemOpen
  \bibfield  {author} {\bibinfo {author} {\bibfnamefont {A.~P.~O.}\
  \bibnamefont {Chan}}, \bibinfo {author} {\bibfnamefont {T.}~\bibnamefont
  {Kvorning}}, \bibinfo {author} {\bibfnamefont {S.}~\bibnamefont {Ryu}}, \
  and\ \bibinfo {author} {\bibfnamefont {E.}~\bibnamefont {Fradkin}},\ }\href
  {\doibase 10.1103/PhysRevB.93.155122} {\bibfield  {journal} {\bibinfo
  {journal} {Phys. Rev. B}\ }\textbf {\bibinfo {volume} {93}},\ \bibinfo
  {pages} {155122} (\bibinfo {year} {2016})}\BibitemShut {NoStop}%
\bibitem [{\citenamefont {Lucas}\ \emph {et~al.}(2016)\citenamefont {Lucas},
  \citenamefont {Davison},\ and\ \citenamefont {Sachdev}}]{Lucas23082016}%
  \BibitemOpen
  \bibfield  {author} {\bibinfo {author} {\bibfnamefont {A.}~\bibnamefont
  {Lucas}}, \bibinfo {author} {\bibfnamefont {R.~A.}\ \bibnamefont {Davison}},
  \ and\ \bibinfo {author} {\bibfnamefont {S.}~\bibnamefont {Sachdev}},\ }\href
  {\doibase 10.1073/pnas.1608881113} {\bibfield  {journal} {\bibinfo  {journal}
  {Proceedings of the National Academy of Sciences}\ }\textbf {\bibinfo
  {volume} {113}},\ \bibinfo {pages} {9463} (\bibinfo {year}
  {2016})}\BibitemShut {NoStop}%
\bibitem [{\citenamefont {Moll}\ \emph {et~al.}(2016)\citenamefont {Moll},
  \citenamefont {Kushwaha}, \citenamefont {Nandi}, \citenamefont {Schmidt},\
  and\ \citenamefont {Mackenzie}}]{Moll1061}%
  \BibitemOpen
  \bibfield  {author} {\bibinfo {author} {\bibfnamefont {P.~J.~W.}\
  \bibnamefont {Moll}}, \bibinfo {author} {\bibfnamefont {P.}~\bibnamefont
  {Kushwaha}}, \bibinfo {author} {\bibfnamefont {N.}~\bibnamefont {Nandi}},
  \bibinfo {author} {\bibfnamefont {B.}~\bibnamefont {Schmidt}}, \ and\
  \bibinfo {author} {\bibfnamefont {A.~P.}\ \bibnamefont {Mackenzie}},\ }\href
  {\doibase 10.1126/science.aac8385} {\bibfield  {journal} {\bibinfo  {journal}
  {Science}\ }\textbf {\bibinfo {volume} {351}},\ \bibinfo {pages} {1061}
  (\bibinfo {year} {2016})}\BibitemShut {NoStop}%
\end{thebibliography}%

\end{document}